\documentclass[aps,prx,twocolumn,superscriptaddress,showpacs,floatfix]{revtex4-1}
\usepackage{amsmath,amssymb,graphicx}
\usepackage{wasysym}
\usepackage[utf8]{inputenc}
\usepackage[T1]{fontenc}
\usepackage{xcolor}
\allowdisplaybreaks
\usepackage{rotating} 
\usepackage[columnwise]{lineno}

\usepackage{textcomp}

\usepackage{bm}

\IfFileExists{siunitx.sty}{\usepackage{booktabs,siunitx}}{}

\usepackage{color}
\definecolor{LinkColor}{rgb}{0.256,0.439,0.588}
\usepackage{hyperref}
\hypersetup{
    pdfstartview={FitH},    
    colorlinks=true,       
    linkcolor=blue,          
    citecolor=blue,        
    filecolor=magenta,      
    urlcolor=blue           
}

\usepackage{pifont}

\newcommand{\La}{\line (1,0  ){12}}
\newcommand{\Lb}{\line (3,5 ){6}}

\newcommand{\Ld}{\line (-1,0){12}}
\newcommand{\Le}{\line (-3,-5){6}}

\newcommand{\C} {\circle*{4}}

\newcommand{\LaT}{\rule[-1pt]{0.4cm}{0.2em}}  
\newcommand{\LdT}{\rule[-1pt]{0.4cm}{0.2em}}  
\newcommand{\LbT}{\rotatebox{60}{\rule[-1pt]{0.4cm}{0.2em}}}  
\newcommand{\LeT}{\rotatebox{60}{\rule[-1pt]{0.4cm}{0.2em}}}  

\newcommand{\pA}{\put(-6,-10)}
\newcommand{\pB}{\put(6,-10)}
\newcommand{\pC}{\put(12,0)}

\newcommand{\pZ}{\put(0,0)}

\newcommand{\pAT}{\put(-6.8,-10)} 
\newcommand{\pBT}{\put(5.2,-10)}  

\newcommand{\rhomb}{
  \pA{\C}\pB{\C}\pZ{\C}\pC{\C}
 }

\newcommand{\rhombH}{
  \begin{picture}(22,10)(-8,-6)
    \pA{\LaT}\pB{\Lb}\pZ{\Le}\pZ{\LdT}
    \rhomb
  \end{picture}
}
\newcommand{\rhombV}{
  \begin{picture}(22,10)(-8,-6)
   \pA{\La}\pBT{\LbT}\pAT{\LeT}\pC{\Ld}
    \rhomb
  \end{picture}
}

\begin{document}

\title{Hidden orders and phase transitions for the fully packed quantum loop model on the triangular lattice
}

\author{Xiaoxue Ran}
\affiliation{Department of Physics and HKU-UCAS Joint Institute of Theoretical and Computational Physics,The University of Hong Kong, Pokfulam Road, Hong Kong SAR, China}

\author{Zheng Yan}
\email{zhengyan@westlake.edu.cn}
\affiliation{Department of Physics, School of Science and Research Center for Industries of the Future, Westlake University, Hangzhou 310030,  China}
\affiliation{Institute of Natural Sciences, Westlake Institute for Advanced Study, Hangzhou 310024, China}

\author{Yan-Cheng Wang}
\affiliation{Hangzhou International Innovation Institute, Beihang University, Hangzhou 311115, China}

\author{Rhine Samajdar}
\affiliation{Department of Physics, Princeton University, Princeton, NJ 08542, USA}

\author{Junchen Rong}
\affiliation{Institut des Hautes \'Etudes Scientifiques, 91440 Bures-sur-Yvette, France}

\author{Subir Sachdev}
\affiliation{Department of Physics, Harvard University, Cambridge, MA 02138, USA}
\affiliation{School of Natural Sciences, Institute for Advanced Study, Princeton, NJ 08540, USA}

\author{Yang Qi}
\email{qiyang@fudan.edu.cn}
\affiliation{State Key Laboratory of Surface Physics, Fudan University, Shanghai 200433, China}
\affiliation{Center for Field Theory
	and Particle Physics, Department of Physics, Fudan University,	Shanghai 200433, China}
\affiliation{Collaborative Innovation Center of Advanced Microstructures, Nanjing 210093, China}

\author{Zi Yang Meng}
\email{zymeng@hku.hk}
\affiliation{Department of Physics and HKU-UCAS Joint Institute of Theoretical and Computational Physics,The University of Hong Kong, Pokfulam Road, Hong Kong SAR, China}

\begin{abstract}
\noindent{\bf Abstract} \\
Quantum loop and dimer models are prototypical correlated systems with local constraints, which are not only intimately connected to lattice gauge theories and topological orders but are also widely applicable to the broad research areas of quantum materials and quantum simulation. Employing our sweeping cluster quantum Monte Carlo algorithm, we reveal the complete phase diagram of the triangular-lattice fully packed quantum loop model. Apart from the known lattice nematic (LN) solid and the even $\mathbb{Z}_2$ quantum spin liquid (QSL) phases, we discover a hidden vison plaquette (VP) phase, which had been overlooked and misinterpreted as a QSL for more than a decade. Moreover, the VP-to-QSL continuous transition belongs to the $(2$\,$+$\,$1)$D cubic$^*$ universality class, which offers a lattice realization of the (fractionalized) cubic fixed point that had long been considered as irrelevant towards the  O($3$) symmetry until corrected recently by conformal bootstrap calculations. Our results are therefore of relevance to recent developments in both experiments and theory, and facilitate further investigations of hidden phases and transitions.
\end{abstract}


\maketitle


\noindent{\bf Introduction}
\label{sec:intro}\\
In quantum many-body systems, topologically ordered phases~\cite{RS91,Wen1991,RJSS91,Wen2019} and their fractionalized excitations~\cite{Wen1991,Kitaev1997} have attracted much attention because they hold the promise of providing solutions to many mysteries of highly entangled quantum matter as encountered in unconventional superconductors~\cite{Kivelson1987,Baskaran1988,Rokhsar1988}, quantum moir\'e materials~\cite{Andrei2020,Kennes2021}, frustrated quantum magnets~\cite{FengZL17,WenXG17,Feng2018claringbullite,YCWang2018,JJWen2019,YiZhou2017,GYSun2018,Broholm2020,ZZ2020string}, and---more recently---programmable quantum simulators~\cite{Samajdar:2020hsw,Verresen:2020dmk,Semeghini21,Yue2021Order,zhou2022u1,Samajdar2022,myerson2022construction,Verresen2022Unifying,ZYan2022,yanEmergent2023,Cheng_2023Variational}. Concomitantly, the quantum phase transitions between topologically ordered phases and conventional symmetry-breaking states have also been actively investigated with a variety of theoretical and numerical approaches~\cite{Balents2002,Isakov2012,Krishanu2015,Plat2015,YCWang2018,GYSun2018,YCWang2021NC,Yan2021,ZYan2022,Samajdar2022}.

One of the most widely studied models to realize such phase transitions, in either numerical simulations or cold-atom experiments, is the quantum dimer model (QDM) ~\cite{Kivelson1987,Rokhsar1988,SSMV99}. In QDMs, $\mathbb{Z}_{2}$ topologically ordered phases originate from the famous Rokhsar-Kivelson (RK) point~\cite{Rokhsar1988,ZZ2021RK,Moessner2001l,Moessner2001b} and acquire extended phase space on nonbipartite triangular and kagome lattices~\cite{Moessner2001l,Ivanov2004,Ralko2005,Misguich2008,Yan2021}. 
However, away from the RK point, the overall structures of the phase diagrams on various 2D geometries---either bipartite or nonbipartite---such as square and triangular lattices, are still under intensive investigation~\cite{Vernay2006,Ralko2008,Krishanu2015, Plat2015, ZY2021mixed,Yan2021,ZY2022,zhou2022u1}. This is due to the lack of controlled methodology to solve these strongly correlated lattice models with local constraints, and discoveries that either confirm or defy earlier nonrigorous arguments.

\begin{figure*}[t]
	\centering
	\includegraphics[width=\linewidth]{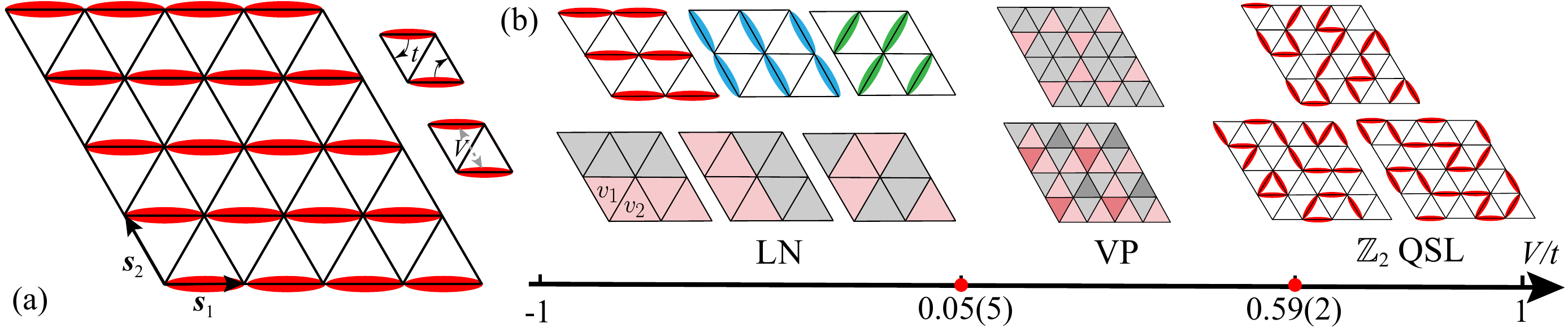}
	\caption{\textbf{Fully packed quantum loop model on the triangular lattice.} (a) Schematic representation of the QLM; $\mathbf{s}_{1}$ and $\mathbf{s}_{2}$ are the primitive vectors. The $t$ and $V$ terms in the Hamiltonian \eqref{eq:eq1} are depicted in the upper-right insets. The dimer configuration sketched is one of the three LN patterns, with fully packed loops along the $\mathbf{s}_{1}$ direction. (b) Phase diagram of the QLM obtained from our simulations. The first row of the left subfigure illustrates the three LN dimer configurations, corresponding to the three vison patterns shown on the second row. The triangles $v_1$ and $v_2$ represent two sublattices for the visons, and the red and grey colors in the triangles denote vison numbers $(\pm1)$ with opposite sign. The first-order phase transition between the LN and VP states occurs at $V=0.05(5)$. The first row of the middle subfigure is the kinetic energy correlation pattern and the second row is the vison plaquette (VP) pattern based on QMC simulation results, respectively (see Fig.~S3(a,b) in Supplementary Note 3 for values in each triangle). VP is a \textit{hidden} dimer solid phase without dimer order. The right subfigures illustrate the representative dimer coverings in the $\mathbb{Z}_{2}$ QSL phase. The continuous phase transition between the VP phase and the QSL occurs at {$V_{c}=0.59(2)$}.}
	\label{fig:fig1}
\end{figure*}

A QDM where two dimers touch every lattice site is also dubbed the fully packed quantum loop model (QLM)~\cite{Bloete1994,Shannon2004,Jaubert2011,Krishanu2015,Plat2015,ranFully2023}.
In the strongly interacting limit, the QLM shares the Hilbert space and low-energy properties of a hard-core boson model (or spin-1/$2$ XXZ model) on a kagome lattice at $1/3$ filling and many other similar frustrated models with local constraints~\cite{Krishanu2015,Plat2015,Wang2017,XFZhang2018}. 
The QLM is also related to the resonant-valence-loop phase in frustrated spin-1 models~\cite{HongYao2010,WeiLi2014}, in analogy to the relation between the one-dimer-per-site QDM and spin-$\frac12$ models. The triangular-lattice QLM has been investigated by field-theoretical analyses and various numerical approaches~\cite{Krishanu2015,Plat2015}, but, so far, no consistent phase diagram has been obtained. 

At the RK point, the model realizes a $\mathbb Z_2$ quantum spin liquid (QSL) phase, which has even $\mathbb Z_2$ topological order and hosts fractionalized quasiparticles called visons.
Tuning away from the RK point, the model also realizes various topologically trivial symmetry-breaking phases. The transitions from the QSL to the topologically trivial phases can be understood as condensation of the visons. Previous numerical studies~\cite{Krishanu2015,Plat2015} show that decreasing the repulsion between parallel dimers leads to a lattice nematic (LN) solid phase.
Furthermore, the phase transition between the QSL and the LN phase, described by vison condensation, is suggested to have an emergent O($3$) symmetry and therefore, belongs to the 3D O($3$)$^*$ universality class; the $^*$ conveys that the transition is between the O(3) symmetry-breaking phase and a $\mathbb Z_2$ topologically ordered phase (rather than a trivial symmetric phase). 

These two works [Refs.~\cite{Krishanu2015,Plat2015}] employ different methods, namely, exact diagonalization (ED) of small systems and density-matrix renormalization group (DMRG) calculations~\cite{Krishanu2015}, or variational quantum Monte Carlo (QMC) simulations with intermediate system sizes of $12\times12$~\cite{Plat2015}. The numerical evidence presented in these two works, however, disagree on the boundaries of the phase diagram. In particular, the former indicates that the LN--QSL transition occurs at $V/t \sim -0.3$ whereas the latter estimated this point to be at $V/t \sim 0.7$ and found that the LN order is vanishingly small between $V/t \in (0.1, 0.7)$. As we will explain below, this is because there is a hidden ordered phase between the LN solid and the QSL.

Theoretically, the speculation of the LN--QSL transition being in the O($3$)$^*$ universality class is based on the \textit{assumption} that the cubic anisotropy is irrelevant at the $3$D O($3$) Wilson-Fisher fixed point~\cite{aharony1973critical,PhysRevB.84.125136,Adzhemyan:2019gvv,Aharony:2022ajv,Pelissetto:2000ek}.
However, this assumption has been disproven by recent conformal bootstrap analyses~\cite{Chester2021,rong2023o3} demonstrating that the $3$D O($3$) fixed point is unstable and will flow towards the nearby corner cubic fixed point, whose symmetry-breaking phase is the corner cubic phase.
We will show later that there indeed emerge cubic anisotropies in this model which change the O($3$) criticality suggested in previous works to a cubic one. The cubic phase transition occurs between the hidden order and the QSL.

Moreover, a previous study~\cite{Wang2017} on the honeycomb-lattice transverse-field Ising model (which is related to the current QLM by representing the vison excitations with Ising spins) shows a first-order transition instead of a continuous one, with clear face-centered cubic anisotropy at the transition point. Therefore, resolving the controversy in the phase diagram of the triangular-lattice QLM requires an additional and consistent understanding that unifies the aforementioned recent developments~\cite{Krishanu2015,Plat2015,Wang2017,Chester2021}, and provides the correct mechanism of the vison-condensation transition from the $\mathbb Z_2$ QSL to the solid phases. The results in Ref.~\cite{Plat2015} and, in particular, the vanishing of the LN order between $V/t\in (-0.3,0.7)$, already hints at such a reconciliation. This first-order phase transition can also be clearly understood in our work, as we will show below, in that it occurs between the LN and the hidden-order phases.

In summary, our work addresses these longstanding questions and presents a comprehensive picture of the quantum phases and phase transitions of the QLM. Using both QMC and ED analyses, we find that between the proposed LN phase~\cite{Mulder2010} and the $\mathbb{Z}_{2}$ QSL phase, there exists a hidden vison plaquette (VP) solid state without apparent dimer order. The phase transition between the LN and VP phases is first-order,  and that from the VP to the $\mathbb{Z}_{2}$ QSL phase is continuous and belongs to the (2+1)D cubic$^*$ universality class. Moreover, our numerical discoveries of the LN--VP first-order transition and the VP--QSL continuous transition provide a unified understanding that is fully consistent with the recent conformal bootstrap analysis of the relation between the O($3$) and cubic fixed points~\cite{aharony1973critical,PhysRevB.84.125136,Adzhemyan:2019gvv,Aharony:2022ajv,Pelissetto:2000ek} in ($2+1$)D~\cite{Chester2021}. 
Our results for the QLM on the triangular lattice provide a nontrivial realization of this scenario in a setting that is experimentally accessible in programmable quantum simulators based on highly tunable Rydberg atom arrays~\cite{Bernien17,keesling2019quantum,PhysRevLett.124.103601,Ebadi.2021,scholl2021quantum,ZYan2022,PhysRevB.105.174417,Samajdar:2020hsw,Verresen:2020dmk}.

\vspace{\baselineskip}
\noindent{\bf Results}\\
{\noindent\bf The constrained model}
\label{sec:spinons}\\
The Hamiltonian of the QLM on a triangular lattice is defined as
\begin{eqnarray}
  H=&-t&\sum_\alpha \left(
  \left|\rhombV\right>\left<\rhombH\right| + \mathrm{h.c.}
  \right) \nonumber \\
  &+V&\sum_\alpha\left(
  \left|\rhombV\right>\left<\rhombV\right|+\left|\rhombH\right>\left<\rhombH\right|
  \right),
\label{eq:eq1}
\end{eqnarray}
where $\alpha$ represents the three possible orientations of all plaquettes of the triangular lattice, as shown in Fig.~\ref{fig:fig1}(a) for an example of the LN state. The local constraint of our model requires two dimers to touch every triangular-lattice site in any configuration, thereby forming the fully packed quantum loops~\cite{Plat2015}. The kinetic term is controlled by $t$, which changes the dimer covering of every flippable plaquette while respecting the local constraint, and $V$ is the repulsion ($V>0$) or attraction ($V<0$) between dimers facing each other. The special RK point is located at $t=V$ and has an exact $\mathbb{Z}_2$ QSL solution~\cite{Moessner2001l}. We set $t=1$ as the unit of energy in our simulations.
\begin{figure}[tb]
	\centering
	\includegraphics[width=1\columnwidth]{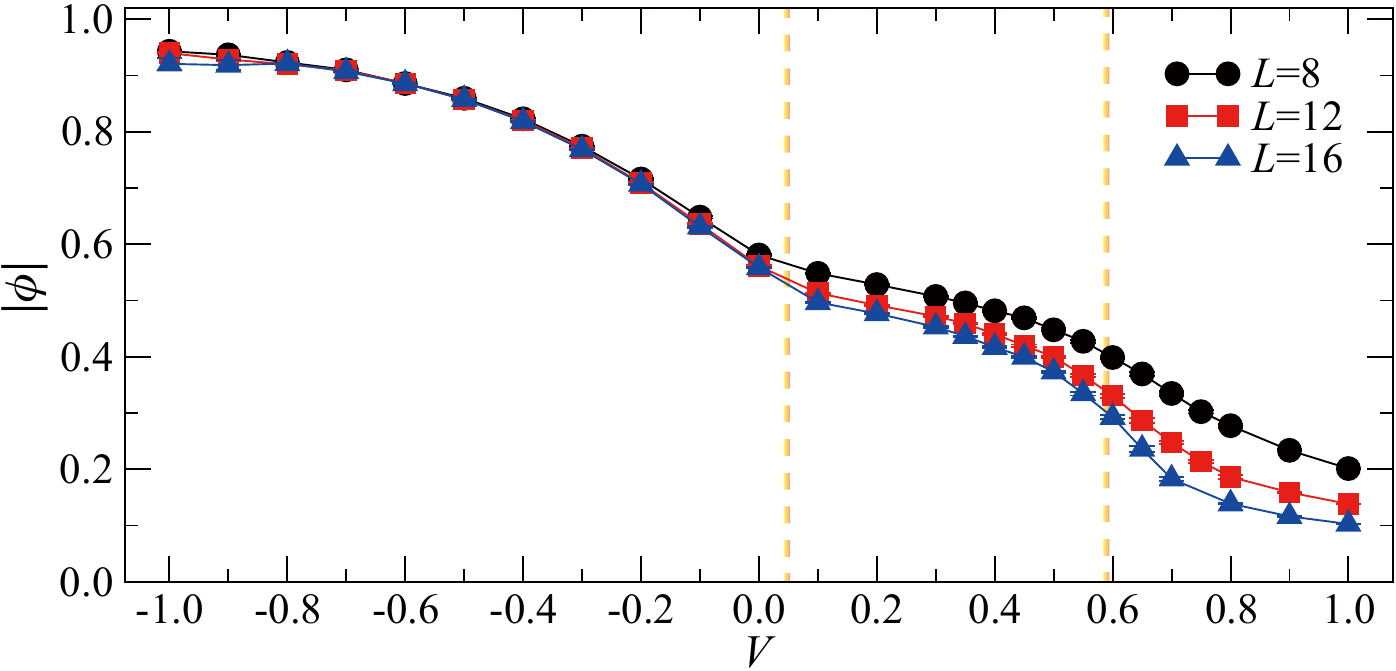}
	\caption{\textbf{Vison O($3$) order parameter $|\bm{\phi}|$ as a function of $V$.} The two dashed lines are guides to the eye for the position of the LN--VP first-order transition and the VP--QSL continuous transition. The error bars here represent the standard error of the mean, which is calculated as $\sigma/ \sqrt{N}$, where $\sigma$ is one standard deviation and $N$ is the total number of independent samples. }
	\label{fig:fig2}
\end{figure}

We employ the recently developed powerful sweeping cluster quantum Monte Carlo method~\cite{Yan2019,ZY2020improved,Yan2021,ZYan2022,ZY2022} to solve this model. Our simulations are performed on the triangular lattice with periodic boundary conditions and system sizes $N=3L^2$ for linear dimensions $L=8,10,12,16,20$, while setting the inverse temperature $\beta=L$. Further description of the QMC implementation and its advantages can be found in Supplementary Note 4.
\vspace{\baselineskip}\\
{\noindent\bf Physical observables}
\label{sec:chargons}\\
To explore the phase diagram of the Hamiltonian in \eqref{eq:eq1}, we first measure the order parameter describing the soft vison modes~\cite{Blankschtein1984,Blankschtein19842,Huh2011,Krishanu2015}, which is given by
\begin{equation}
\phi^{}_{j}=\sum_{\textbf{r}}(v^{}_{1,\textbf{r}},v^{}_{2,\textbf{r}})\cdot\textbf{u}^{}_{j}e^{i \mathbf{M}_j\cdot \textbf{r}}, \quad j=1,2,3,
\label{eq:eq2}
\end{equation}
where $\textbf{r}$ runs over all the unit cells of the triangular lattice, as shown in Fig.~\ref{fig:fig1}. The vector $\bm{\phi}$\,$=$\,$(\phi_{1},\phi_2,\phi_3)$ encapsulates the 3D order parameters of the visons, which are obtained from the Fourier transforms of vison configurations $ v_i v_j $\,$=$\,$(-1)^{N_{P_{ij}}}$, with $N_{P_{ij}}$ being the number of dimers cut along the path $P$ between triangular plaquettes $i$ and $j$~\cite{Yan2021}. 
The label $(v_{1,\textbf{r}},v_{2,\textbf{r}})$ denotes visons living at the centers of the two sublattices of the triangular plaquette at $\textbf{r}$ [see the left subfigure in Fig.~\ref{fig:fig1}(b)]. In our simulations, we choose a reference vison on a certain triangular plaquette, e.g., by setting $v_{1,\textbf{r}=(0,0)}$\,$=$\,$\pm 1$ on the upper triangle of the first unit cell. The vison configurations are obtained with respect to such a reference, from which we can assign values of $(v_{1,\textbf{r}},v_{2,\textbf{r}})$ for all $\textbf{r}$. The three momenta $\mathbf{M}_{1}$\,$=$\,$(\pi/\sqrt{3},\pi/3)$, $\mathbf{M}_{2}$\,$=$\,$(\pi/\sqrt{3},-\pi/3)$, and $\mathbf{M}_{3}=(0,2\pi/3)$ correspond to the
low-energy modes of the vison dispersion and are invariant under the projective symmetry transformations~\cite{Huh2011,Krishanu2015} as we show in Supplementary Note 1. The associated $\mathbf{u}_{j}$ are $(1,\ 1)^T$ for $\mathbf{M}_{1}$ and $\mathbf{M}_{2}$, and $(1,\ -1)^T$ for $\mathbf{M}_{3}$. The derivation of \eqref{eq:eq2}, the symmetry analysis of the 3D order parameter of the visons, and representative vison configurations in the VP phase are presented in the Supplementary Note 1.

According to the analysis of the effective critical theory\cite{Huh2011,Krishanu2015}, the low-energy effective Hamiltonian of the quantum loop model is dual to the ferromagnetic transverse field Ising model on the honeycomb lattice. In the large-external-field limit, the spin in the $z$ direction fluctuates, representing the fluctuation of the boson number on each site, which corresponds to the $\mathbb{Z}_{2}$ spin liquid. As the Ising coupling increases, the energy dispersion acquires its minima at three $M$ points, which correspond to three patterns of the LN phase. Therefore, the transition between the $\mathbb{Z}_{2}$ QSL and the nontopological phase can be regarded as the process of vison condensation, which can be described by the vison order parameter $(\phi_{1},\phi_{2},\phi_3)$\cite{Krishanu2015,Wang2017}. We take the magnitude $|\bm{\phi}|=\sqrt{\phi_{1}^{2}+\phi_{2}^{2}+\phi_{3}^{2}}$, plotted in Fig.~\ref{fig:fig2}, as the order parameter to roughly detect the solid phases and their transitions to the $\mathbb{Z}_2$ QSL phase. 
As illustrated in Fig.~\ref{fig:fig2}, the order parameter $|\bm{\phi}|$ clearly vanishes via a two-step process that sharpens with increasing system sizes. When $V<0$, the order parameter is finite, denoting long-range order in the vison pattern corresponding to the LN phase, as shown in the left subfigure in Fig.~\ref{fig:fig1}(b); when $V>0$, $\phi$ drops to a finite plateau, signifying another symmetry-breaking phase. 
It is worth noting that the order vanishes around $V=0$ if we measure it through dimer correlations (e.g., see Fig.S5 of SI). Ww thus ask: why does the nature of the order seem so different when viewed in terms of dimers and visons? 
\begin{figure*}[t]
	\centering
	\includegraphics[width=\linewidth]{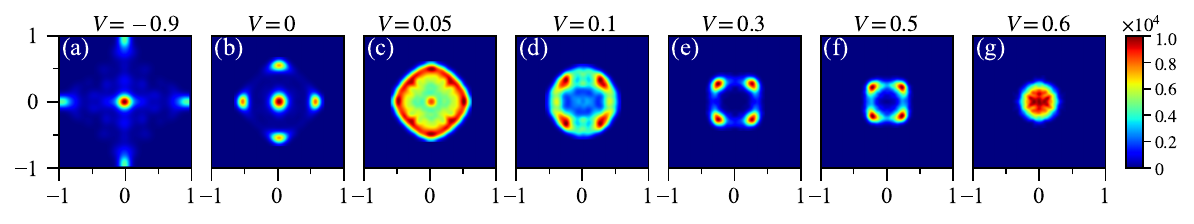}
	\caption{\textbf{Histograms of the O($3$) order parameter.} The histograms are plotted on the two-dimensional ($\phi_1$,$\phi_2$) plane and are obtained from the QMC data for $L = 24$ systems with different $V$. The (a,b) face-cubic anisotropies are observed inside the LN phase, while the histogram in a region of phase coexistence, plotted in (c), indicates the first-order nature of the transition. The (d, e, f) corner-cubic anisotropies are observed inside the VP phase, and (g) near the continuous phase transition point.}
	\label{fig:fig3}
\end{figure*}

To further reveal the nature of the symmetry-breaking phases, we plot the histogram of the order parameter $\bm\phi$ in Fig.~\ref{fig:fig3}.
The distribution of $\bm\phi$ is indeed different in the two phases:
it is peaked at the six face centers of the cube in the phase for $V<0$, and at the eight corners of the cube in the phase for $0<V<0.6$.
According to the symmetry transformation of $\bm\phi$ in Eqs.(S8) and (S9) of the SI,
a peak at the face center corresponds to breaking threefold rotational symmetry while preserving translational and twofold rotational symmetries. From the broken symmetries, we recognise that this is the LN phase.
On the other hand, a peak at the corner is associated with breaking the translational symmetry in a manner that doubles the unit cell in both directions while preserving the rotational symmetries.
Therefore, the ground state actually belongs to a \textit{plaquette-ordered} phase with a $2$\,$\times$\,$2$ unit cell, which we refer to as the VP phase.
As we will discuss later, the ground state possesses emergent hidden order in that it does not exhibit any symmetry breaking in the dimer-dimer correlations but does so in other correlation functions. 
The LN and VP ordered phases break incompatible symmetries and thus cannot be connected via spontaneous symmetry breaking, which points to a first-order phase transition between them. The histogram in a region of phase coexistence, plotted in Fig.~\ref{fig:fig3} (c), also indicates the first-order nature of the transition. More detailed results on the evolution from the LN to the VP solid phase can be seen in Supplementary Note 2.

\vspace{\baselineskip}
{\noindent\bf Hidden order: the vison plaquette phase}\\
\label{hidden order}
It is only natural to ask why the vison plaquette phase had not been identified for more than a decade~\cite{Krishanu2015, Plat2015}. This is likely because, except for the vison order parameter histogram in Fig.~\ref{fig:fig3}, one cannot find a corresponding dimer order in the VP phase, i.e., the order is {\it hidden} in the dimer basis.
In order to uncover this hidden order in the VP phase, we perform the following analysis.

Motivated by the QMC data, we use exact diagonalization (ED) to further demonstrate that the hidden order cannot be measured in the dimer basis. As there is no clear peak in the structure factor of dimer correlations in the VP phase for all the system sizes simulated, we seek assistance from the vison histograms to amplify any signals of possible symmetry-breaking phases in the dimer patterns. In the VP phase, the eight cubic fixed points stay in eight octants in the histograms [such as in Figs.~\ref{fig:fig3}(d), (e), and (f)], so every vison configuration can be classified into a certain octant except those on the boundaries; since two vison classes in opposite octants correspond to the same dimer configuration, there are only four classes of dimer configurations. We classify QMC dimer configurations into these four classes and average the ones in the same class (which would amplify the symmetry breaking in the dimers, if there is any) to obtain the dimer density on the strongest vison-ordered configurations. We collected about $500,000$ ($300,000$) such configurations for an $L=4$ ($L=8$) system at $V=0.4$ but found that the real-space dimer density is homogeneous, $\sim 1/3$, on all the bonds of the lattice.

Furthermore, as shown in 
 Supplementary Note 6 (and Fig.S4), a similar analysis using the ground-state wavefunction from ED of a $4\times4$ lattice at $V=0.3$ (which eliminates the statistical error in QMC simulations) also confirms that there is no translational symmetry breaking in the dimer density distribution.
Therefore, despite evidence of translational symmetry breaking in Fig.~\ref{fig:fig3}, both QMC and ED results strongly suggest that the VP order is a hidden-order phase in the dimer basis, with a homogeneous dimer occupation.
We note that such a unique hidden symmetry-breaking phase has not been reported earlier in the QDM literature. Whether there are intricate (emergent) symmetries that actually protects the homogeneity is an interesting question for future investigation.

This unsuccessful detection in the dimer basis can also be understood easily from the vison configurations. In fact, from the real-space vison correlation function in Fig.~S3
(b) of Supplementary Note 3, the difference in correlation functions between two closest triangles is seen to be a constant (about $0.14$ at $V=0.3$). It is well known that this difference can be translated to the dimer occupation on the bond separating two neighboring triangles~\cite{MoessnerSondhi2001b}.
Thus, the constant difference in vison correlations also points to a uniform dimer occupation. It is worth emphasizing that a homogeneous dimer density only requires a constant difference between closest visons instead of a constant vison density. This gives rise to the possibility that this VP order exists separately from the QSL.

Considering that the vison is a fractional quasipaticle which cannot be measured in experiments while the QDM and QLM are widely investigated in frustrated quantum materials and blockaded ultracold atomic arrays, we now propose a measurable observable to identify this hidden order.
We notice that the translational symmetry breaking indicated by the histograms in Fig.~\ref{fig:fig3}(d), (e) and (f) is indeed reflected in other correlation functions.
Firstly, this can be seen from the vison correlations $\langle v_iv_j\rangle$, as displayed in the phase diagram in Fig.~\ref{fig:fig1} and discussed in detail in Supplementary Note 3.
We find that although the vison correlation function appears to change sign under mirror reflections and sixfold rotations, the physical state actually preserves these symmetries, because of the two-to-one correspondence between vison and dimer configurations.
Furthermore, the translational symmetry breaking can also be detected from correlation functions of local operators.
\begin{figure}[tb]
	\centering
	\includegraphics[width=1\columnwidth]{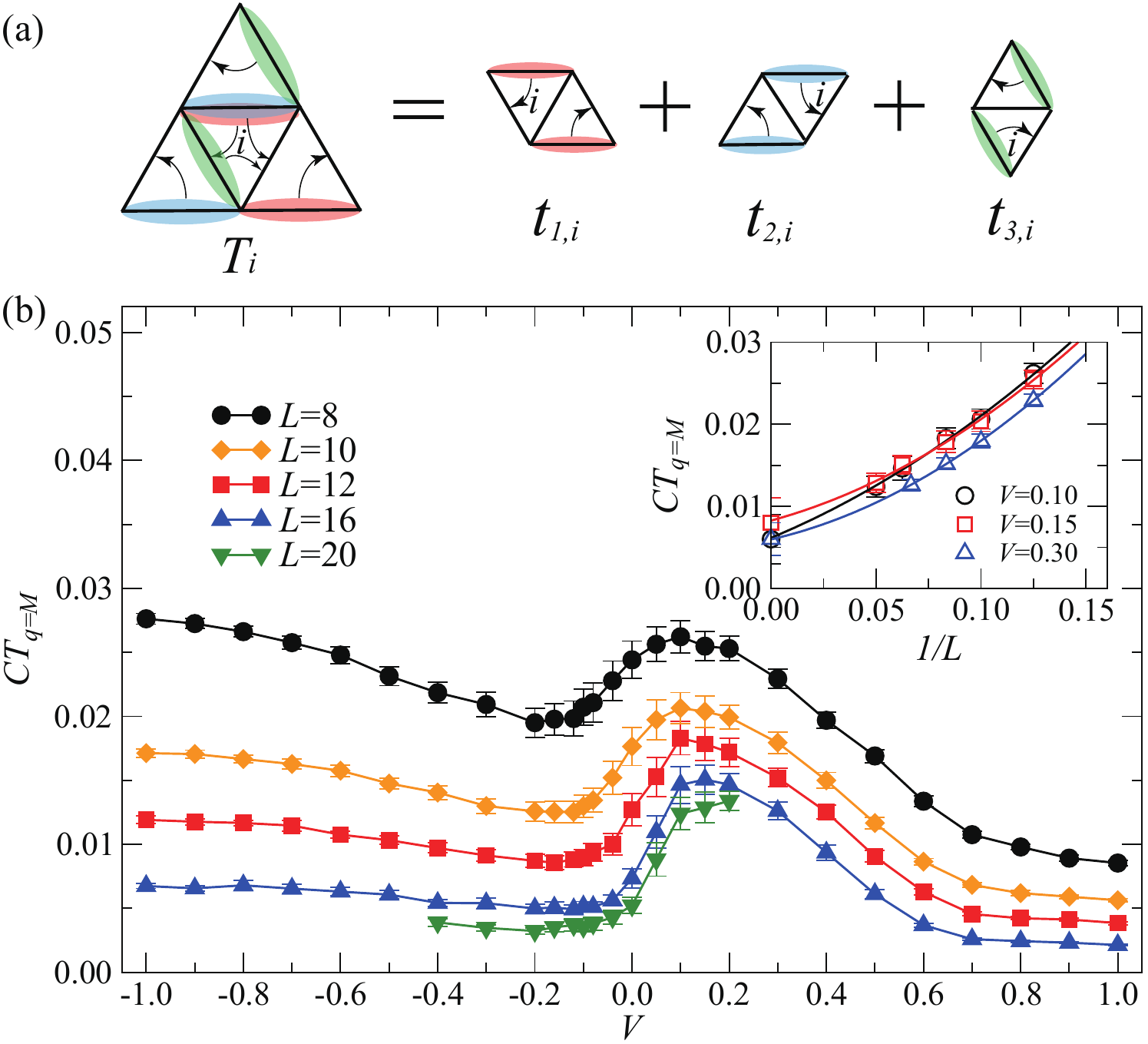}
	\caption{ \textbf{Hidden order in the VP phase.} (a) The definition of the operator $T_i$ acting on triangle $i$. Each operator $T_i$ includes the $t$-terms of the Hamiltonian~\eqref{eq:eq1} on the three rhombi with labels $1,2,3$ enclosing the triangle $i$. (b) The structure factor at the $M$ point, $CT_{\mathbf{q}=M}$, as a function of $V$; in the VP phase, $CT_{\mathbf{q}=M}$ acquires long-range order. The inset shows that the VP phase (for $V=0.1,0.15,$ and $0.3$, the extrapolated value at $L=\infty$ is $0.006(2), 0.008(2),$ and $0.0097(3)$, respectively) persists in the thermodynamic limit. The error bars here represent the standard error of the mean, which is calculated as $\sigma/\sqrt{N}$, where $\sigma$ is one standard deviation and $N$ is the total number of independent samples.}
	\label{fig:fig5}
\end{figure}

Therefore, we construct the following composite order parameter which can be measured in experiments:
\begin{equation}
    \label{eq:tpo}
    T^{}_i=t^{}_{1,i}+t^{}_{2,i}+t^{}_{3,i},
\end{equation}
where the sum runs over the kinetic $t$-terms of the Hamiltonian on three rhombi with labels $1,2,3$ containing the triangle $i$ [Fig.~\ref{fig:fig5}(a)].
This is a natural choice of a composite order parameter because it preserves the threefold rotational symmetry.
The correlation function of $CT$\,$\equiv$\,$\langle T_i T_j \rangle$ shows a structure-factor peak at the $\mathbf{M}$ point of momentum space in the VP phase, as shown in Fig.~\ref{fig:fig5}(b). The corresponding real-space $CT$ correlation function---plotted in Fig.S3
(a) and sketched in the middle subfigure of Fig.~\ref{fig:fig1}(b)---clearly reveals the pattern of translational symmetry breaking and the associated $2\times2$ unit cell. 

We can also try to identify $CT_{\mathbf{q}=\mathbf{M}}$ with operators of the low-energy effective action \eqref{eq:eq3}. The operator should break translational symmetry while preserving the sixfold rotational symmetry. 
Since $CT_{\mathbf{q}=\mathbf{M}}$ is constructed from dimers in \eqref{eq:tpo}, it should also be gauge invariant. The natural candidate is therefore $X=\phi_1\phi_2+\phi_2\phi_3-\phi_3\phi_1$, which is clearly invariant under the sixfold rotation operation defined in Eq. (S9) of Supplementary Note 1. 
After this identification, we can infer the critical behavior of $CT_{\mathbf{q}=\mathbf{M}}$ near the second-order phase transition at $V\sim 0.6$  from the knowledge of the scaling dimension of $X$ of the cubic conformal field theory (CFT). 
As mentioned previously, the cubic CFT and the O($3$) CFT have similar critical exponents; since $X$ belongs to the $j=2$ representation of O($3$) (the T representation in the notation of Ref.~\cite{Chester2021}), we obtain the critical exponent $\eta^{*}\approx 1.42$ \cite{Chester2021,PhysRevB.84.125136,rong2023o3}, which is close to the critical exponent for the (2+1)D O(2)$^{*}$ transition observed before in a frustrated kagome magnet~\cite{Isakov2012,YCWang2018}.
\begin{figure*}[!]
	\centering
	\includegraphics[width=0.8\textwidth]{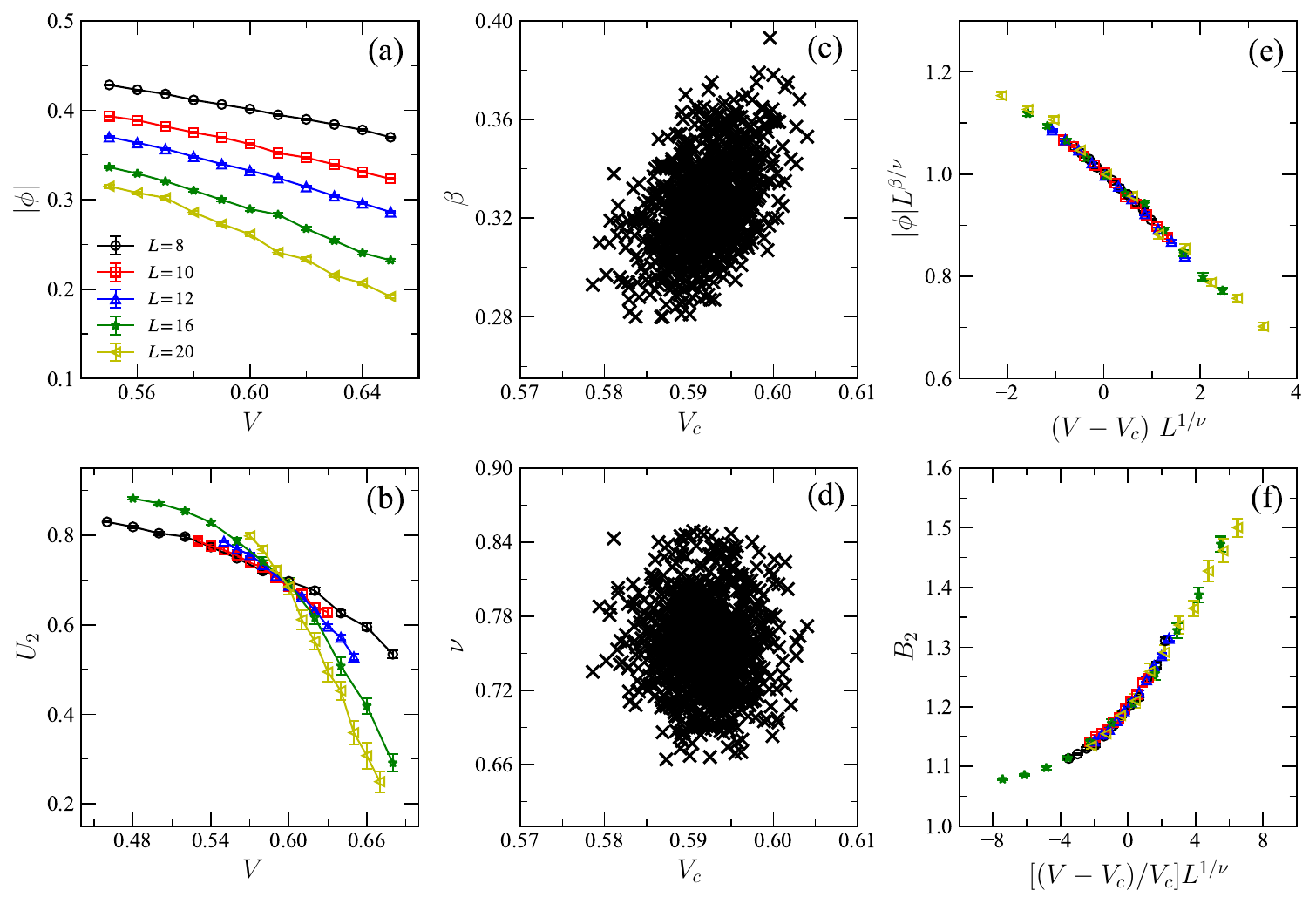}
	\caption{\textbf{The cubic fixed point.}  Stochastic data collapse with both the vison order parameter $|\phi|$ and its Binder cumulant. (a) and (b) plot the vison order parameter and its Binder cumulant  $U_2$. All data points are connected with polynomial fits to the third order and $\chi^2/\mathrm{d.o.f}$ ($\chi^2$ per degree of freedom) is close to 1. The location of the crossing in (b) shifts due to finite-size effects. (c) and (d) are the distributions of $\beta$ and $\nu$ with $V_c$ obtained from the bootstrap sampling process. Each data point is determined from a fitted curve of the data collapse for $|\phi|$ and $B_2$, and the the $\chi^2/\mathrm{d.o.f}$ of all the curves is close to 1. We obtain $V_c=0.59(2)$, $\beta=0.33(5)$, and $\nu=0.75(8)$. The values of $\nu$ and $\beta$ are consistent with the standard O(3) value, $\nu=0.7112(5)$ and $\beta=0.3689(3)$. (e) and (f) show the data collapse of the vison order parameter and its Binder ratio  $B_2$. Here, the values of $V_c$, $\beta$, and $\nu$ are determined independently from our fitting procedures. The error bars here represent the standard error of the mean, which is calculated as $\sigma/\sqrt{N}$, where $\sigma$ is one standard deviation and $N$ is the total number of independent samples.}
	\label{fig:fig4}
\end{figure*}

\vspace{\baselineskip}
{\noindent\bf The cubic$^*$ phase transition}\\
\label{mean-field1}
From the histograms in Fig.~\ref{fig:fig3} (e), (f) and (g), we note that there is a sizeable cubic anisotropy order parameter even close to the phase transition point between the VP and QSL phases. Theoretically, the cubic anisotropy was thought to be irrelevant at the $3$D O($3$) Wilson-Fisher fixed point~\cite{aharony1973critical,PhysRevB.84.125136,Adzhemyan:2019gvv,Aharony:2022ajv,Pelissetto:2000ek}.
However, recent conformal bootstrap analyses~\cite{Chester2021,rong2023o3} demonstrate that the $3$D O($3$) fixed point is unstable and will flow towards the nearby cubic fixed point, whose symmetry-breaking phase is the corner cubic phase. Thus, the phase transition point here is indeed of the cubic$^*$ criticality, with the $*$ representing that the order parameter here is constructed by a fractional quasiparticle. 

We now turn to the critical behavior of the VP--QSL transition shown in Fig.~\ref{fig:fig4}. From the plot of the vison order parameter $|\bm{\phi}|$ as a function of $V$ in Fig.~\ref{fig:fig2} for different system sizes, the transition around $V=0.6$ is clearly continuous, and---per the discussion above---belongs to the cubic$^{*}$ CFT. However, the critical exponents of the cubic fixed point and those of the O($3$) fixed point are not practically distinguishable with the resolution in our study (the scaling dimension of the cubic anisotropy differs by only $\sim 0.01$~\cite{Chester2021}). 
For example, the most recent classical Monte Carlo simulations give $-0.00001<\nu_{O(3)}-\nu_{cubic}<0.00007$ and $\eta_{O(3)}-\eta_{cubic}=-0.00061(10)$~\cite{Hasenbusch:2022zur}, confirming an earlier prediction of a perturbative calculation ~\cite{PhysRevB.67.024418}. 
The conformal bootstrap result in Ref.~\cite{Chester2021} also implies that the difference should be $\sim 10^{-4}$.
Thus, the logic is that if we find a cubic anisotropy of the order parameter near the phase transition and the critical exponents are the same as the O(3) criticality up to QMC precision, then this amounts to the cubic phase transition being demonstrated numerically.

 Therefore, in Fig.~\ref{fig:fig4}, we perform stochastic data collapse with both the vison order parameter $\phi$ and its Binder cumulant~\cite{beach2005data,wang2006high} as shown in Figs.~\ref{fig:fig4}(a) and (b); the details can be found in Supplementary Note 7. By including more data points and a larger parameter space for the optimization process, our scheme, as shown in Figs.~\ref{fig:fig4}(b) and (e), gives independent estimates of $V_c=0.59(2)$, $\beta=0.33(5)$, and $\nu=0.75(8)$, which are consistent with the O$(3)$ critical exponents of $\nu=0.7112(5)$ and $\beta=0.3689(3)$. Furthermore, in Figs.~\ref{fig:fig4}(e) and (f), $\phi$ and its Binder ratio are illustrated with a rescaled $x$-axis to show the high-quality data collapse. Our numerical analyses firmly establish that the QLM in \eqref{eq:eq1} realizes different limits of the effective action (\eqref{eq:eq3} in the next section) and are consistent with the latest understanding of the fixed points of cubic symmetry. 

\vspace{\baselineskip}
{\noindent\bf Effective action and renormalization-group analysis}\\
\label{sec:2min}
In order to fully understand the LN--VP first-order transition and the VP--QSL continuous transition, as well as their fundamental relation to the O($3$) and cubic fixed points in $(2+1)$D, 
we will need to first explain the effective action and the renormalization-group (RG) analysis of the problem in a more general and up-to-date setting. The low-energy description of these transitions~\cite{Krishanu2015}, with the O($3$) order parameter $\bm{\phi}$ in \eqref{eq:eq2}, can be cast into the action $S = \int dt d^2x \, L$ with the Lagrangian
\begin{align}
&L=  \sum_{i=1}^3(\partial^{}_{\mu} \phi^{}_i )^2+ r \sum_{i=1}^3 \phi_i^2 + \mu \left(\sum_{i=1}^3 \phi^{}_i\phi^{}_i\right)^2 +\nu^{}_4 \sum_{i=1}^3 \phi_i^4\nonumber\\
&+\dots,
\label{eq:eq3}
\end{align}
coupled to a $\mathbb{Z}_2$ gauge theory. The $\ldots$ denote higher-order terms, whose explicit forms can be found in the SI. The $\mathbb{Z}_2$ gauge theory does not change the dynamics of the system; its effect is to gauge out operators that are odd under the $\mathbb{Z}_2$ symmetry which flips the sign of all three $\phi_i$ fields. The first three terms in \eqref{eq:eq3} preserve the O($3$) symmetry. The $\nu_4$ piece (ignoring other higher-order terms), on the other hand, breaks the O(3) symmetry to a three-dimensional cubic symmetry. This cubic model has a long history since its first appearance in describing the structural phase transition of perovskites~\cite{aharony1973critical,Aharony:2022ajv} (for a review, see Ref.~\cite{Pelissetto:2000ek}).
More recently, the experimentally observed structural phase transitions of single-layer transition metal dichalcogenides~\cite{duerloo2014structural,li2016structural,wang2017structural}, such as MoS$_2$, have also been described by the effective action above~\cite{PhysRevMaterials.2.114002}.

Mean-field theory suggests that there are two phases: when $\nu_4>0$, in the symmetry-broken phase ($r<0$), the minima of the effective potential are located at
\begin{equation}
\langle\phi_1\rangle=\pm v, \quad \langle\phi_2\rangle=\pm v,\quad  \langle\phi_3\rangle=\pm v.
\end{equation}
Such a phase is commonly called the corner-cubic phase (i.e., the VP phase in our case), yielding the bright spots in the histograms in Figs.~\ref{fig:fig3}(d)-(f); here, the eight possible states are in one-to-one correspondence with the eight vertices of a three-dimensional cube.  When
$\nu_4<0$, the symmetry-broken phase is described by
\begin{equation}
\langle\phi_1\rangle=\pm v',\quad \langle\phi_2\rangle=\langle\phi_3\rangle=0,
\end{equation}
together with the other states where either $\langle\phi_2\rangle$ or $\langle\phi_3\rangle$ acquires an expectation value. The six symmetry-broken states in this case correspond to the face centers of a three-dimensional cube, thus defining a face-cubic phase  (i.e., the LN phase in our case which corresponds to the bright spots in the histograms in Figs.~\ref{fig:fig3}(a) and (b)], which has also been seen in the honeycomb-lattice transverse-field Ising model~\cite{Wang2017} that bears a low-energy action similar to \eqref{eq:eq3}.

\begin{figure}[tb]
	\centering
	\includegraphics[width=1\columnwidth]{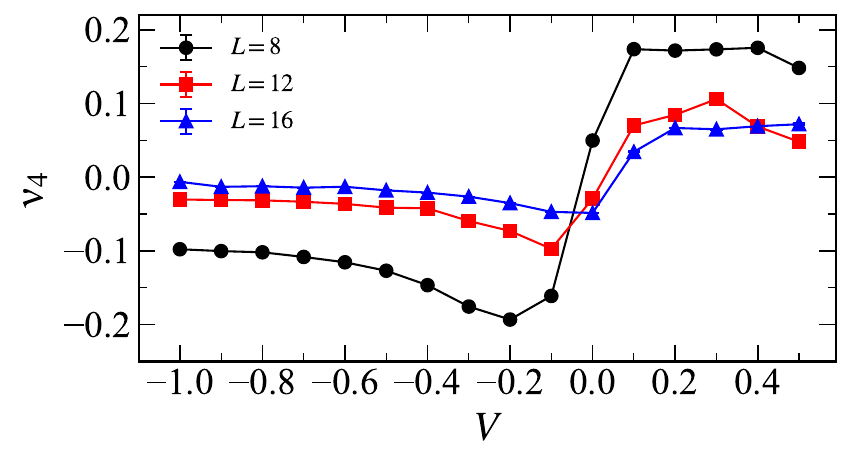}
	\caption{\textbf{Evidence for the first-order phase transition.} The figure shows the coefficient of the anisotropy $\nu_4$ in the effective action~\eqref{eq:eq3} as function of $V$, which is computed from the QMC histogram data of the O($3$) order parameter in Fig.~\ref{fig:fig3}. $\nu_4$ changes sign at the LN--VP transition, signifying the change from face- to corner-cubic anisotropy in the effective action of the lattice model. The error bars here represent the standard error of the mean, which is calculated as $\sigma/\sqrt{N}$, where $\sigma$ is one standard deviation and $N$ is the total number of independent samples.}
	\label{fig:fig6}
\end{figure}

To further demonstrate the first-order phase transition here, with the above background, we now move back to the QMC data in Figs.~\ref{fig:fig3} and \ref{fig:fig6}. 
It is important to notice that at the cubic fixed point, the coupling constant $\nu_4$ is a small positive number. This leaves room for another weakly first-order phase transition when $\nu_4$ changes sign. This is precisely the transition between the corner- and face-cubic phases that we observed at $V/t\approx0.05$ in the vison order-parameter histograms in Fig.~\ref{fig:fig3}.

To quantitatively describe the phase transition between the LN and VP phases, we consider the anisotropy parameter $\nu_{4}$ in Eq.\eqref{eq:eq3}. In the Monte Carlo simulations, it can be expressed as~\cite{Wang2017} (see derivations in Supplementary Note 5)
\begin{align}
\nu_{4} &= -\frac{1}{vol}\frac{15\sqrt{\pi}}{\langle \phi^{4}\rangle}\left(\langle Y^{0}_{4}\rangle+\frac{\langle \phi^2 Y^{0}_{4}\rangle\langle \phi^4\rangle\langle \phi^6\rangle-\langle Y^{0}_{4}\rangle\langle\phi^6\rangle^2}{\langle \phi^6\rangle^2-\langle \phi^4\rangle\langle \phi^8\rangle}\right),
\end{align}
where $Y^{0}_{4}$ and $Y^{0}_{6}$ are two spherical harmonics given by $Y^{0}_{4} = 3(3-30\cos^{2}{\theta}+35\cos^{4}{\theta})/(16\sqrt{\pi})$,
$Y^{0}_{6} = \sqrt{13}(-5+105\cos^{2}{\theta}-315\cos^{4}{\theta}+231\cos^{6}{\theta})/(32\sqrt{\pi})$, and $\theta$ is the polar angle of the order parameter $\bm{\phi}=(\phi_1, \phi_2, \phi_3)$ which is defined in \eqref{eq:eq2} for every component. The volume factor is $vol=L^2\beta$. Notice the leading term in the above equation essentially measures the angular dependence of the expectation of $\langle\phi(\theta,\psi)\rangle$, projecting onto $Y^{0}_{4}$, with the signs of this term different in the corner- and face- cubic phases. This term gives the leading contribution to $\nu_4$. As shown in Figs.~\ref{fig:fig6}, the sign of the anisotropy parameter $\nu_{4}$  changes from negative to positive near the LN--VP transition. A similar approximate emergent continuous symmetry at a first-order transition has been reported in the checkerboard $J$-$Q$ model in both 2D and 3D lattices~\cite{BWZhao2018,GYSun2021}; the latter is in fact related to ongoing experimental efforts in understanding the thermodynamic data of the Shastry-Sutherland material SrCu$_2$(BO$_3$)$_2$ under high pressure~\cite{Zayed17,JGuo2020,Jimenez2021}.

This analysis also tells us that the phase transition from the disordered phase to the corner-cubic phase is second-order and continuous because the RG flow suggests that when $\nu_4>0$, the theory~\eqref{eq:eq3} flows to a conformal field theory.
It remains to answer the question to which universality class the $\nu_4>0$ phase transition belongs. This seemingly easy question was under debate for many years; see Ref.~\cite{Aharony:2022ajv} for a review of early theoretical studies. It was not until much later that a consistent answer was obtained using three different methods: lattice Monte Carlo simulations \cite{PhysRevB.84.125136}, perturbative field-theory calculations \cite{Adzhemyan:2019gvv},
and the conformal bootstrap \cite{Chester2021}. It turns out that the continuous phase transition at $\nu_4>0$ is in a new universality class, which is different from the O($3$)-invariant Heisenberg model.
The question of the universality class is closely related to the scaling dimension of the cubic anisotropy operator $\mathcal{O}=\sum_{i=1}^3 (\phi_i)^4$.
Among the above-mentioned methods, the conformal bootstrap approach \cite{rattazzi2008bounding,poland2019conformal} provides a concrete nonperturbative proof that $\Delta_\mathcal{O} <3$ \cite{Chester2021,rong2023o3}.
This indicates that the O($3$) conformal fixed point is unstable against cubic anisotropy, and therefore, the phase transition should be in a new universality class. 

We denote the corresponding CFT the cubic fixed point. Since, in our case, the symmetric phase is a $\mathbb Z_2$ QSL with fractionalized vison excitations, the VP--QSL transition is actually of the cubic$^*$ universality, with the same critical exponents $\beta$ and $\nu$ as the cubic CFT, but it acquires a large anomalous dimension $\eta^{*}$ since the order parameter is actually made of a composite object of the fractionalized particles. Such $^*$-type transitions between a symmetry-breaking phase and a $\mathbb Z_2$ QSL have been shown in kagome-lattice frustrated magnetic models with (2+1)D O(2)$^{*}$ universality and $\eta^{*} \sim 1.5$~\cite{Isakov2012,YCWang2021NC,YCWang2018}.

\vspace{\baselineskip}
\noindent{\bf Relation to experiments on Rydberg atom arrays}\\
\label{sec:continuum_more}
Recently, Ref.~\cite{Samajdar:2020hsw} demonstrated that the phases of various triangular-lattice quantum dimer models can be realized using Rydberg atoms arrayed on the sites of a kagome lattice via programmable optical tweezers. In particular, this presents a direct route to the experimental realization of the fully packed quantum loop model studied in this paper in quantum simulators based on interacting Rydberg atoms. In such a setup, strong 
van der Waals interactions prevent the simultaneous excitation of neighboring atoms to the Rydberg state in a phenomenon known as the Rydberg blockade. This Rydberg blockade competes with the
chemical potential (the laser detuning) which favors or disfavors occupation of the Rydberg states. 

Such a blockaded \textit{kagome}-lattice Rydberg array can be mapped to a dimer model on the \textit{triangular} lattice with two dimers per site~\cite{Samajdar:2020hsw} in certain parameter regimes. Numerical studies \cite{Samajdar:2020hsw,ZYan2022} have in fact suggested that both the LN and QSL phases of the triangular-lattice QLM may arise in this setup. Therefore, it would be interesting to investigate the possible existence of the hidden-order VP phase discovered in this study in the experimental atomic system.

Importantly, this hidden order also bears significant implications for the experimental identification of $\mathbb{Z}_2$ QSLs in Rydberg atom arrays. In recent works~\cite{Verresen:2020dmk,ZYan2022,Semeghini21}, a diagonal loop operator has been used to detect the $\mathbb{Z}_2$ QSL phase in a model of Rydberg atoms with emergent gauge-charged Ising matter degrees of freedom \cite{Samajdar2022}. This operator is defined as an arbitrary loop that runs across several bond centers, i.e., $Z\equiv(-1)^{c}$, where $c$ counts the number of dimers cut by the loop. 
While our model has no Ising matter, such a loop operator will have a nonzero expectation value in the VP phase as well, \textit{even though} it has no topological order but only a hidden translation symmetry breaking. Thus, the hidden VP phase actually reveals that the nonzero diagonal loop operator plus a disordered diagonal density is not a sufficient condition for a QSL.
\vspace{\baselineskip}
\noindent{\bf Discussion}\\
\label{sec:disc}
In this work, using the newly developed sweeping cluster QMC algorithm, supplemented with ED and symmetry analysis of the vison order parameter, we mapped out the entire phase diagram of the triangular-lattice QLM in an unbiased manner. Besides the lattice nematic and $\mathbb Z_2$ quantum spin liquid phases, we discover a hidden vison plaquette phase via off-diagonal measurements in the dimer basis. This VP phase is invisible to diagonal measurements and obeys the local constraint of a $\mathbb{Z}_2$ gauge field. Our results reveal that the LN--VP first-order transition is triggered by the change from face- to corner-cubic anisotropy of the O($3$) vison order parameter, and the VP--QSL continuous transition, driven by the condensation of visons, is of the cubic$^*$ universality class---which is very close to the $(2+1)$D O($3$) one---in full consistency with recent conformal bootstrap findings on the cubic fixed point~\cite{Chester2021,rong2023o3}.

The VP solid phase discovered here exhibits clear translational symmetry breaking in the vison correlation functions and in the dimer hopping order in Fig.~\ref{fig:fig5} but has no apparent dimer density order in our ED and QMC simulations. That is why it had been treated as a part of the QSL phase for a long time. It represents a hidden state of quantum matter and, at the same time, resolves the previous controversy regarding the phase boundary between the QSL and LN phases in Refs.~\cite{Krishanu2015, Plat2015} (note that Ref.~\cite{Plat2015} had observed the vanishing dimer order parameter characteristic of this phase in a parameter regime consistent with ours).

In light of recent experiments with programmable quantum simulators based on highly tunable Rydberg atom arrays, our results could also have direct experimental relevance. The experiment in Ref.~\cite{Semeghini21} probes the case where the atoms are positioned on the \textit{links} of the kagome lattice, connecting to the quantum dimer model on the kagome lattice~\cite{Verresen:2020dmk}. Our study on the triangular-lattice quantum loop model connects to the case where the atoms are placed on the \textit{sites} of the kagome lattice~\cite{Samajdar:2020hsw}. Investigation of the LN, hidden-order VP, and $\mathbb{Z}_2$ QSL phases---as well as their subtle phase transitions in the context of the O($3$) and cubic fixed points---can inspire new experiments.
In particular, for the case with Rydberg atoms on the kagome sites, an analog of the VP order is a promising possibility for the region proximate to an even QSL~\cite{Samajdar2022} or the so-called string phase identified in  Ref.~\cite{Samajdar:2020hsw}.
Finally, the rich interplay between the VP and QSL phases is a promising avenue for related future numerical and theoretical studies of this  system~\cite{ZYan2022,Samajdar2022}.

\vspace{\baselineskip}
\noindent{\bf Method}\\
{\noindent\bf \textbf{Sweeping cluster algorithm.}} \\
This is a  quantum Monte Carlo method developed by the authors which can work well in constrained spin models~\cite{Yan2019,ZY2020improved,Yan2021,ZYan2022,ZY2022}. The key idea of the sweeping cluster algorithm is to sweep and update layer by layer along the imaginary time direction, so that the local constraints (gauge field) are recorded by update-lines. In this way, all the samplings are done in the restricted Hilbert space, i.e., the low-energy space. In this article, we can measure the information about single visons because in a strictly constrained space, the energy gap of other quasiparticles, such as spinons, becomes infinitely large and thus these quasiparticles do not exist in the restricted Hilbert space. Recently, this method has even been developed for simulating systems with soft constraints~\cite{ZYan2022} or higher-order constraints~\cite{ZY2022}.

\vspace{\baselineskip}
\noindent{\bf Data availability}

The data that support the findings of this study are available from the authors upon reasonable request.

\vspace{\baselineskip}
\noindent{\bf Code availability}

All numerical codes in this paper are available upon reasonable request to the authors.

\vspace{\baselineskip}
\noindent{\bf References}
\bibliographystyle{longapsrev4-2}
\bibliography{main}

\vspace{\baselineskip}
\noindent{\bf Acknowledgements}\\
We thank Ning Su for insightful discussions on the O($3$) and cubic fixed points, and thank Subhro Bhattacharjee,  Frank Pollmann, Fabian Alet, and Sylvain Capponi for valuable discussions on the phase diagram of the QLM over the years. XR, ZY and ZYM acknowledge support from the Research Grants
Council of Hong Kong SAR of China (Project Nos. 17301420, 17301721, AoE/P-701/20, 17309822, HKU C7037-22G), the ANR/RGC Joint Research Scheme sponsored by RGC of Hong Kong and French National Research Agency (Project No.A\_HKU703/22), the K. C. Wong Education Foundation (Grant No.~GJTD-2020-01), and the Seed Funding ``Quantum-Inspired explainable-AI'' at the HKU-TCL Joint Research Centre for Artificial Intelligence. YQ acknowledges support from the National Natural Science
Foundation of China (Grant Nos. 11874115 and 12174068). JR is supported by Huawei Young Talents Program at IHES. Y.C.W. acknowledges  support from Zhejiang Provincial Natural Science Foundation of China (Grant Nos. LZ23A040003). RS and SS are supported by the U.S. Department of Energy under Grant DE-SC0019030. We thank the Tianhe-II platform at the National Supercomputer Center in Guangzhou, the HPC2021 system under the Information Technology Services and the Blackbody HPC system at the Department of Physics, University of Hong Kong for their technical support and generous allocation of CPU time. We thank the Beijing PARATERA Tech CO., Ltd. (URL: https://cloud.paratera.com).

\vspace{\baselineskip}
\noindent{\bf Author Contributions}

ZY and ZYM initiated this project. XR performed the QMC simulations. ZY developed the Sweeping Cluster QMC method.  YCW did the ED calculations. JR and YQ did the field analysis. RS and SS put forward the experimental measurements for hidden order. All authors contributed to the analysis of the results and writing manuscript. YQ and ZYM supervised the project.

\vspace{\baselineskip}
\noindent{\bf Competing interests}

The authors declare no competing interests.\\




\setcounter{equation}{0}
\setcounter{figure}{0}
\renewcommand{\theequation}{S\arabic{equation}}
\renewcommand{\thefigure}{S\arabic{figure}}

\clearpage
\newpage
\setcounter{page}{1}
\begin{widetext}
\linespread{1.05}

\centerline{\bf Supplemental Material for ``Hidden orders and phase transitions} 
\centerline{\bf  for the fully packed quantum loop model on the triangular lattice"} 
\vskip3mm

\centerline{}
In this {\color{black}Supplementary Information}, we first discuss the derivation of the O($3$) vison order parameter and its symmetry transformations (Supplementary Note 1). Then, Supplementary Note 2 illustrates the histograms of the order parameter obtained from QMC simulations, showing further details of the intricate symmetry-breaking patterns in the LN and VP phases. In the Supplementary Note 3, we  discuss the real-space vison correlation function and the real-space symmetry-breaking pattern of the kinetic term in the VP phase. In the Supplementary Note 4, the implementation of the sweeping cluster quantum Monte Carlo simulation, with the constraint of two dimers per site, is discussed. Thereafter, Supplementary Note 5 and Supplementary Note 6 present, respectively, details on the QMC measurements of the anisotropy parameters, dimer order parameter, and exact diagonalization (ED) results on a $4\times 4$ system with homogeneous dimer density inside the VP phase as well as comparisons thereof to QMC.  Finally, Supplementary Note 7 carries out an unbiased data collapse of the O(3) order parameter and its Binder ratio and an independent determination of the critical exponents of the VP--QSL transition; we find that the exponents are consistent with the O($3$) ones within error bars.

\section*{SUPPLEMENTARY NOTE 1: $\mathrm{O}$($3$) order parameter}
\label{sec:oop}

The vison condensation transition in the quantum loop model on the triangular lattice can be analyzed by mapping it to the ferromagnetic transverse-field Ising model on the dual honeycomb lattice~\cite{Krishanu2015,Wang2017,Yan2021,Huh2011}, where
\begin{equation}
H^{}_\mathrm{dual}=J^{}_1\sum_{\langle I J \rangle}v_{I}^{z}v_{J}^{z}+J^{}_2\sum_{\langle\langle I J \rangle\rangle}v_{I}^{z}v_{J}^{z}+J^{}_3\sum_{\langle\langle\langle I J \rangle\rangle\rangle}v_{I}^{z}v_{J}^{z}-\Gamma\sum_{I}v_{I}^{x}.
\label{eq:h}
\end{equation}
Here, $I$ and $J$ are the sites on the dual lattice while $J_{1,2, 3}$ is the first-, second-, and third-nearest-neighbor Ising coupling. In the large-$\Gamma$ limit, all spins point in the $\hat{x}$ direction. A flipped spin generates a vison excitation, thus $v_{I}^{z}=+1(-1)$ is the vison creation (annihilation) operator. As we increase $J$, the visons acquire a dispersion and eventually condense at its minima. The Fourier transform of the Hamiltonian in Eq.~\eqref{eq:h} can be written as
\begin{equation}
H^{}_\mathrm{dual}=-\sum_{\mathbf{q}}\psi^{\phantom{\dagger}}_{\mathbf{q}}\,J(\mathbf{q})\,\psi_{\mathbf{q}}^{\dagger},
\label{eq:h2}
\end{equation}
where $\psi^{\phantom{\dagger}}_\mathbf{q}\equiv(v_{1q}^{z},\ v_{2q}^{z})$, and $J(\mathbf{q})$ is the interaction matrix, which takes the form of
\begin{equation}      
	J(\mathbf{q})=\left(                
	\begin{array}{cc}  
		a & b  \\  
		b^* & a \\  
	\end{array}
	\right)             
	\label{jq}
\end{equation}
with $J_1=J_2=J_3=1$. $a$ and $b$ in this matrix are defined as
\begin{align}
a &= e^{i\mathbf{q}\cdot \mathbf{s}_1}+e^{-i\mathbf{q}\cdot \mathbf{s}_1}+e^{i\mathbf{q}\cdot \mathbf{s}_2}+e^{-i\mathbf{q}\cdot \mathbf{s}_2}+e^{i\mathbf{q}\cdot (\mathbf{s}_1+\mathbf{s}_2)}+e^{i\mathbf{q}\cdot (\mathbf{s}_1+\mathbf{s}_2)}, \\
b &= 1+e^{i\mathbf{q}\cdot \mathbf{s}_1}+e^{i\mathbf{q}\cdot \mathbf{s}_2}+e^{i\mathbf{q}\cdot (\mathbf{s}_1+\mathbf{s}_2)}+e^{-i\mathbf{q}\cdot (\mathbf{s}_1+\mathbf{s}_2)}+e^{i\mathbf{q}\cdot (\mathbf{s}_1-\mathbf{s}_2)},
\end{align}
where $\mathbf{s}_1=(\sqrt{3}/2,-3/2)$ and $\mathbf{s}_2=(\sqrt{3}/2,3/2)$ are the primitive vectors sketched in Fig.1
(a) of the main text. The minima of the energy dispersion $J(\mathbf{q})$ occur at three momenta $\mathbf{M}_{1}=(\pi/\sqrt{3},\pi/3)$, $\mathbf{M}_{2}=(\pi/\sqrt{3},-\pi/3)$, and $\mathbf{M}_{3}=(0,2\pi/3)$ in the hexagonal Brilliouin zone, as demonstrated in Fig~\ref{fig:sfig1}. The eigenvectors at these momenta are
\begin{equation}
	\label{eq:u}
	\begin{split}
		u_1=u_2=(1,\ 1)^T,\quad
		u_3=(1,\ -1)^T.
	\end{split}
\end{equation}
Therefore, the vison O($3$) order parameter $\bm{\phi}=(\phi_1, \phi_2, \phi_3)$ can be expressed as
\begin{equation}
	\label{eq:phi}
	\begin{split}
		\phi_1=\sum_\mathbf{r}(v_{1,\mathbf{r}}+v_{2,\mathbf{r}})e^{i\pi x},\quad
		\phi_2=\sum_\mathbf{r}(v_{1,\mathbf{r}}+v_{2,\mathbf{r}})e^{i\pi y},\quad
		\phi_3=\sum_\mathbf{r}(v_{1,\mathbf{r}}-v_{2,\mathbf{r}})e^{i\pi (y-x)},
	\end{split}
\end{equation}
where $x$ and $y$ are the coordinates of all plaquettes of the triangular lattice.

\begin{figure}[t]
	\centering
	\includegraphics[width=0.6\columnwidth]{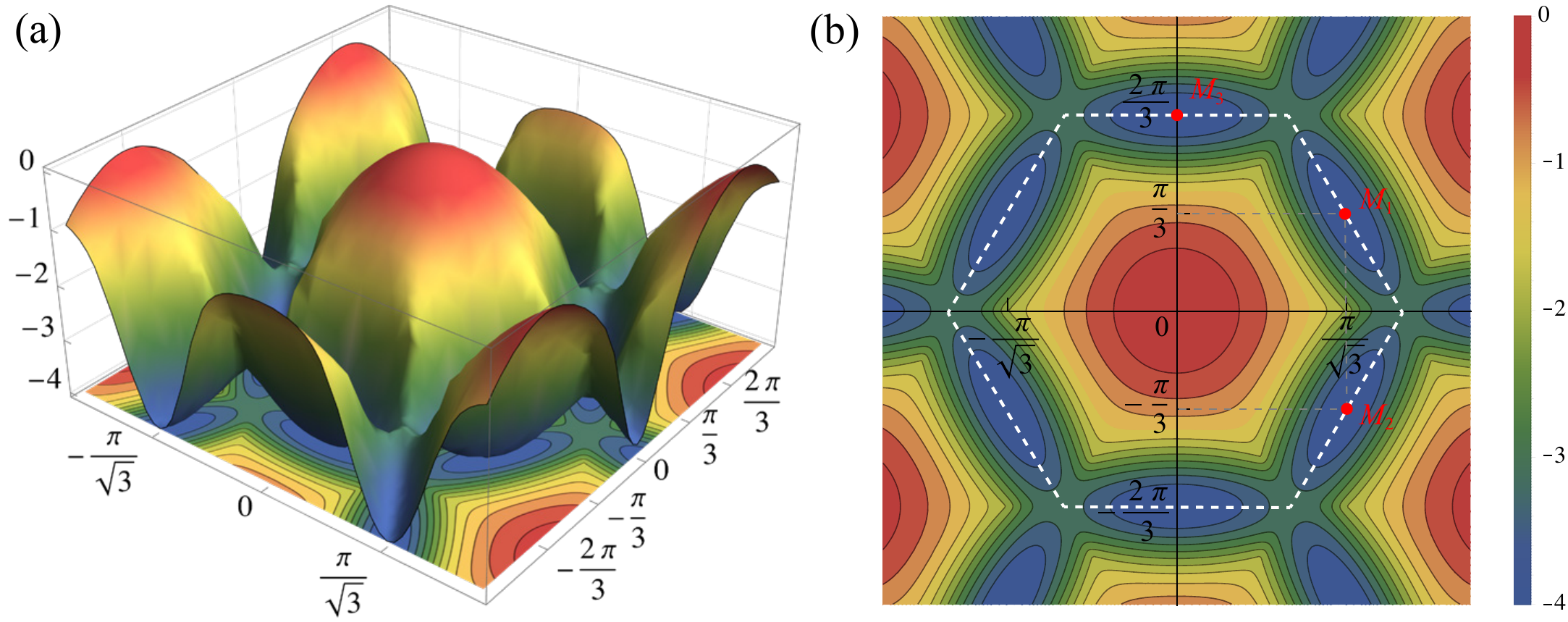}
	\caption{(a) The dispersion of $J(\mathbf{q})$  and (b) the Brilliouin zone (BZ) of the honeycomb lattice. The minima of the energy dispersion occur at $\mathbf{M}_{1}=(\pi/\sqrt{3},\pi/3)$, $\mathbf{M}_{2}=(\pi/\sqrt{3},-\pi/3)$, $\mathbf{M}_{3}=(0,2\pi/3)$ and the three other symmetry-equivalent $\mathbf{M}$ points. }
	\label{fig:sfig1}
\end{figure}

The O($3$) order parameter $\bm{\phi}=(\phi_1,\ \phi_2,\ \phi_3)$ can transform under various symmetries. The transformation matrices for $\bm{\phi}$ with respect to translations ($T_x, T_y$), bond inversion ($\mathcal{I}$), rotation ($R_6$) , global $Z_2$, and mirror symmetry  along the $y$ axis ($\mathcal{M}$), are given by
\begin{alignat}{5}
	T_x&=\left(
	\begin{array}{ccc}
		-1 & 0 & 0 \\
		0 & 1 &  0\\
		0 &  0 & -1\\
	\end{array}
	\right),\quad
	&&T_y&&=\left(
	\begin{array}{ccc}
		1 & 0 & 0 \\
		0 & -1 &  0\\
		0 &  0 & -1\\
	\end{array}
	\right),\quad
	&&\mathcal{I}&&=\left(
	\begin{array}{ccc}
		1 & 0 & 0 \\
		0 & 1 &  0\\
		0 &  0 & -1\\
	\end{array}
	\right),
	\label{I}\\
	R_6&=\left(
	\begin{array}{ccc}
		0 & 0 & -1 \\
		1 & 0 &  0\\
		0 &  1 & 0\\
	\end{array}
	\right),
	\label{r6}
\quad
	&&Z_2&&=\left(
	\begin{array}{ccc}
		-1 & 0 & 0 \\
		0 &-1 &  0\\
		0 &  0 & -1\\
	\end{array}
	\right),
\quad
	&&\mathcal{M}&&=\left(
	\begin{array}{ccc}
		0 & 0 & -1 \\
		0 &1 &  0\\
		-1 &  0 & 0\\
	\end{array}
	\right).
\end{alignat}
These matrices generate a finite subgroup of O($3$) which contain 48 elements~\cite{Krishanu2015}. This group is isomorphic to $\mathbb{Z}_2\times \mathbb{S}_4$, with $\mathbb{S}_4$ being the permutation group of four elements.

\section*{SUPPLEMENTARY NOTE 2: Order parameter histograms}
\label{sec:histo}

Besides the order parameter histograms presented in the main text, here, we add more detailed results on the evolution from the LN to the VP solid phase. Figure~\ref{fig:sfig2} compiles the histograms for a $L=\beta=16$ system, as $V$ is varied from $-1$ to $1$. It is clear that for cases with $V<0.1$, the histograms exhibit the face-cubic distribution, and for the cases with $V>0.1$, the histograms exhibit the corner-cubic distribution.

\begin{figure}[t]
	\centering
	\includegraphics[width=1.0\columnwidth]{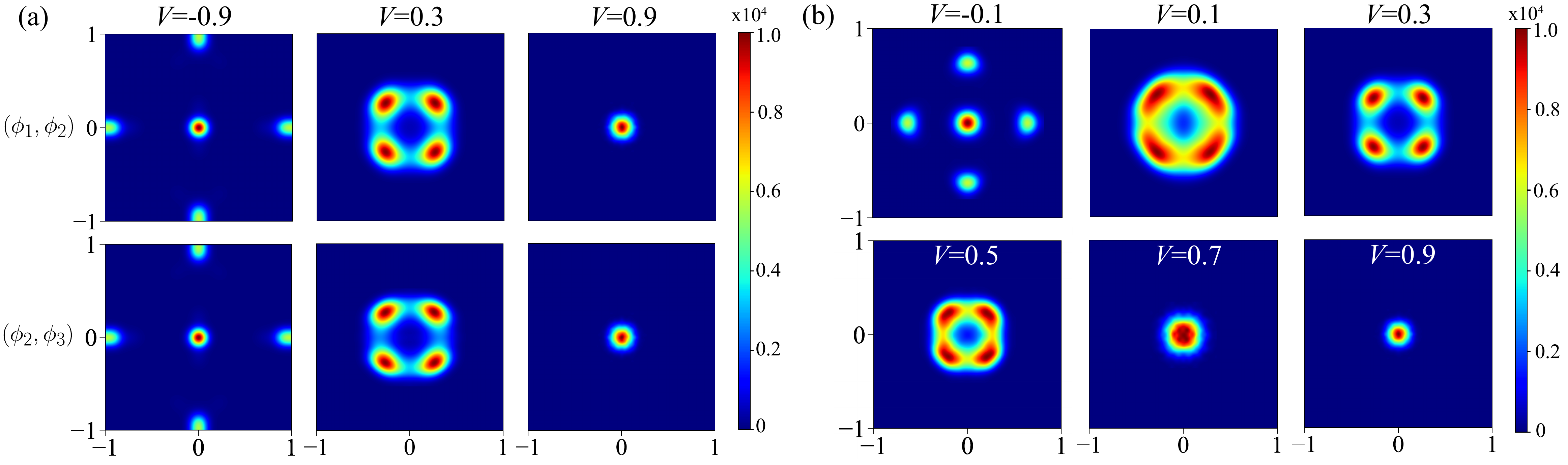}
	\caption{Order parameter histograms for a $L=16$ system. (a) Projection of the O($3$) order parameter on the ($\phi_1,\phi_2$) plane and the ($\phi_2,\phi_3$) plane. The histograms in the three phases (LN with $V=-0.9$, VP with $V=0.3$, and QSL with $V=0.9$) are shown. (b) Evolution of the histogram by tuning $V$, from LN, through VP, to QSL.}
	\label{fig:sfig2}
\end{figure}

Close to the LN-to-VP transition, i.e., the panel with $V=0.1$ in Fig.~\ref{fig:sfig2}(b), the histogram develops asymptotic O($3$) symmetry with very weak corner-cubic anisotropy (this is because $V=0.1$ lies on the VP side of the transition). Inside the VP phase, and as $V$ increases towards the VP--QSL continuous transition point $V_c = 0.598$ (see below), the histograms [in panels with $V=0.3, 0.5$ in Fig.~\ref{fig:sfig2}(b)] begin to shrink in diameter, suggesting the reduction of the vison order parameter. Eventually, inside the QSL phase [panels with $V=0.7,0.9$ in Fig.~\ref{fig:sfig2}(b)], the histogram becomes a point at $(0,0,0)$, indicating that the vison order parameter completely vanishes in the topologically ordered QSL phase.

\section*{SUPPLEMENTARY NOTE 3: Real-space vison correlations}
\label{sec:vison}

To investigate the symmetry-breaking pattern of the VP phase, we calculate the correlations of the operator $T$ described in Eq.~($9$) of the main text. This operator is defined on a triangular plaquette from three related kinetic ($t$) terms, $T_i=t_{1,i}+t_{2,i}+t_{3,i}$, [see inset of Fig.~\ref{fig:sfig3}(a)]. Here, $i$ labels the triangle, and $1,2,3$ denote the three kinds of kinetic ($\sim t$) terms (rhombi) in the Hamiltonian acting on this triangle. The correlation of the operators between the triangular plaquettes $i$ and $j$ is $\langle T_iT_j\rangle$.

We calculate the real-space vison correlations, selecting the leftmost and lowest red triangle below as a reference (its correlation is identically 1),
as illustrated in Fig.~\ref{fig:sfig3}(b). 
Two visons living on the centers of related triangular plaquettes $i$ and $j$ can be connected with a open string; their correlation is $\langle v_i v_j\rangle=\langle(-1)^{N_{P_{ij}}} \rangle$ and ${N_{P_{ij}}}$ is the number of dimers cut along the path $P$ between plaquettes $i$ and $j$. In the two-dimer-per-site case, $\langle v_i v_j \rangle$ is independent on the chosen path, i.e., the visons here are gauge invariant and measurable.

We found that the $T$-operator and vison correlations clearly exhibit the translational symmetry breaking of the VP solid, but preserve $C_6$ rotational symmetry. Although the vison correlation function appears to change sign under mirror-reflection and sixfold rotations, the physical state actually preserves these symmetries, because of the two-to-one correspondence between vison and dimer configurations. Individual triangles with positive (negative) values of vison correlations gather into big triangles. In addition, the central triangle of the thus-formed big triangle always has a larger absolute value, which is about three times that of the other triangles in the same big triangle. The translation periods along the $\mathbf{s}_{1}$ and $\mathbf{s}_{2}$ directions are both $2$, which means that the unit cell is a $2\times 2$ rhomboid.

\begin{figure}[htp]
	\centering
	\includegraphics[width=1\columnwidth]{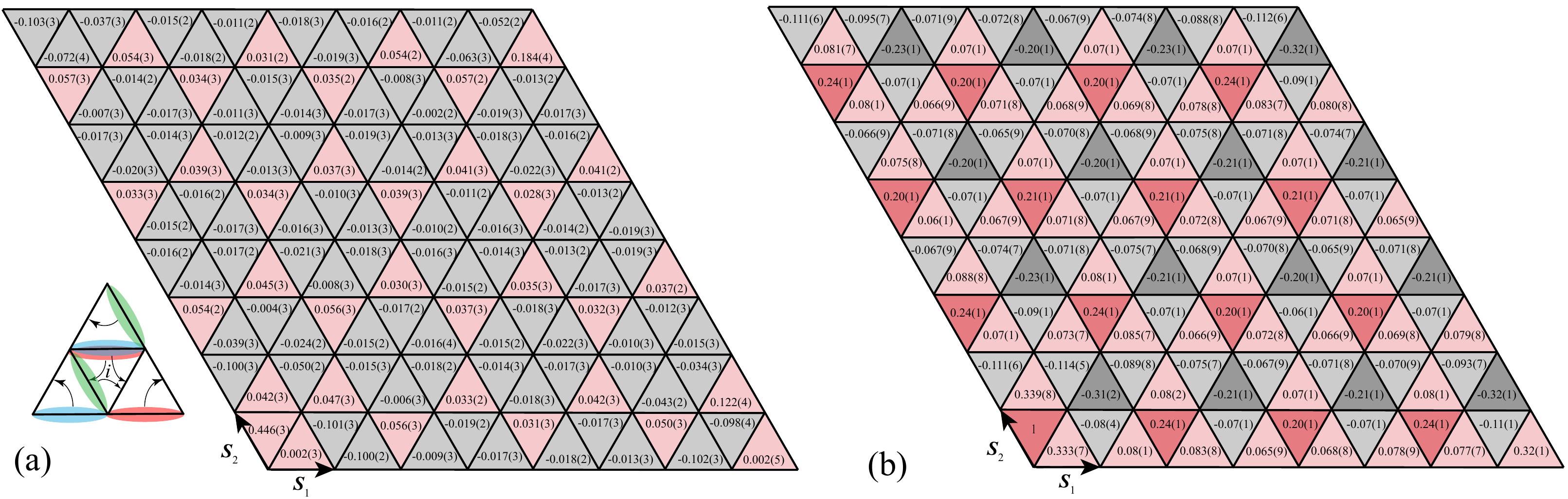}
	\caption{(a) Real-space $T$-operator correlations of the $L=8$ system at $V=0.3$. The inset shows the summation of the three $t$ terms in the Hamiltonian.  (b) Real-space vison correlations of the $L=8$ system at $V=0.3$; $\mathbf{s}_{1}$ and $\mathbf{s}_{2}$ are the primitive vectors. For both figures, we set the  leftmost and lowest triangle as the reference. The red (grey) color conveys the positive (negative) value of the correlation.} 
	\label{fig:sfig3}
\end{figure}

\section*{SUPPLEMENTARY NOTE 4: Sweeping cluster quantum Monte Carlo with two dimers per site}
\label{sec:qmc}

The sweeping cluster QMC approach employed in this work is a new method developed by us, which works well in constrained quantum lattice models~\cite{Yan2019,ZY2020improved,Yan2021,ZYan2022}. 
Prior to the development of sweeping cluster QMC, in order to solve the QDM or QLM types of constrained models, one had to rely on either exact diagonalization of small systems,  variational approaches such as DMRG that suffer from finite-size effects on cylindrical geometries~\cite{Krishanu2015}, projector Monte Carlo approaches (which include the Green's function~\cite{Ivanov2004,Ralko2005,Ralko2006,Vernay2006,Ralko2007,Plat2015} and diffusion Monte Carlo schemes~\cite{OFS2005,OFS2006}), or sampling directly in height space and discarding the unconstrained configurations~\cite{Banerjee_2013,PhysRevB.90.245143,PhysRevB.94.115120}. These projector Monte Carlo methods obey the geometric constraints but are not efficient away from the RK point~\cite{OFS2005walk} and only work at $T=0$. Furthermore, there does not exist any cluster update schemes for the projector methods. On the contrary, the sweeping cluster algorithm, based on the world-line Monte Carlo scheme~\cite{OFS2002,Alet2005a,Alet2005b}, is designed to sweep and update layer-by-layer along the imaginary-time direction so that the local constraints (gauge fields) are recorded by the update lines. In this way, all the samplings are performed in the restricted Hilbert space and it provides a cluster update scheme for constrained systems~\cite{ZY2020improved} that works at all temperatures. Proper finite-size-scaling analyses can then be carried out to explore phase transitions and critical phenomena. 


To implement the constraint of having two dimer per site in our QMC simulation, we set the initial state as one of the three LN patters with the same probability~\cite{Plat2015}. In the region of the LN phase, the equilibrium process will not change the orientation of the initial state; therefore, the randomly selected initial state ensures each pattern of the LN phase has an equal probability of $\frac{1}{3}$  to emerge. In the VP and QSL regions, one can also use such initialization schemes, and we have tested that in these regions too, they produce the same QMC results as a completely random initialization. We simulate $L\times L$ triangular lattices with system sizes $L=8,12,16$ with the inverse temperature set to $\beta=L$ and $10^4$ Monte Carlo samplings were used to obtain average values of the observables in all calculations.

In this study, we can measure the information about single visons because in a strictly constrained space, the energy gap of other quasiparticles such as spinons, becomes infinitely large and thus these quasiparticles do not exist in the restricted Hilbert space; the vison is thus well-defined here.

\section*{SUPPLEMENTARY NOTE 5: Anisotropy parameters}
\label{sec:e}
The effective action of the O($3$) transition is given by,
\begin{align}
S=&\int \mathrm{d}t\, \mathrm{d}x^2 \sum_{i=1}^3(\partial^{}_{\mu} \phi^{}_i )^2+ r \sum_{i=1}^3 \phi_i^2 + \mu \left(\sum_{i=1}^3 \phi^{}_i\phi^{}_i\right)^2 +\nu^{}_4 \sum_{i=1}^3 \phi^{4}_i+\mu^{}_6 \left(\sum_{i=1}^3 \phi_i^2\right)^3 + \nu^{}_6 (\phi^{}_1\phi^{}_2\phi^{}_3)^2+\nu_6' \left(\sum_{i=1}^3 \phi_i^2\right)\left(\sum_{i=1}^3 \phi_i^4\right).
\label{eq:seq8}
\end{align}
Assuming that the coupling constants $\nu_4$, $\nu_6$, $\nu_6' $ are all much smaller than one, the averages of the spherical harmonics are
\begin{align}
\frac{\langle Y^{0}_{4}\rangle}{vol} &= -\frac{1}{15\sqrt{\pi}}\nu^{}_4\langle\phi^4\rangle_0-\frac{1}{15\sqrt{\pi}}\nu_6'\langle\phi^6\rangle_0-\frac{1}{330\sqrt{\pi}}\nu^{}_6\langle\phi^6\rangle_0,\nonumber\\
\frac{\langle \phi^2Y^{0}_{4}\rangle}{vol} &= -\frac{1}{15\sqrt{\pi}}\nu^{}_4\langle\phi^6\rangle_0-\frac{1}{15\sqrt{\pi}}\nu_6'\langle\phi^8\rangle_0-\frac{1}{330\sqrt{\pi}}\nu^{}_6\langle\phi^8\rangle_0,\nonumber\\
\frac{\langle Y^{0}_{6}\rangle}{vol} &= -\frac{1}{231\sqrt{13\pi}}\nu^{}_6\langle\phi^6\rangle_0.
\label{eq:seq9}
\end{align}
Here, $\langle .\rangle_{0}$ denotes the averages under Boltzmann weights determined by $S_0$, where $S_0$ is the action~\eqref{eq:seq8} with the anisotropic couplings $\nu_4$, $\nu_6$ and $\nu_6'$ turned off. The symbol $\phi$ denotes the length of $\phi_i$, i.e., $\phi=(\mu (\sum_{i=1}^3 \phi_i\phi_i)^2)^{1/2}$. The volume factor $vol=L^2\beta$ for lattice system. The two spherical harmonics are
\begin{align}
Y^{0}_{4} &= \frac{3}{16\sqrt(\pi)}(3-30\cos^{2}{\theta}+35\cos^{4}{\theta}), \\
Y^{0}_{6} &= \frac{\sqrt{13}}{32\sqrt{\pi}}(-5+105\cos^{2}{\theta}-315\cos^{4}{\theta}+231\cos^{6}{\theta}).
\end{align}
As an example, we explain here how to derive the formula for $\langle Y^{0}_{4}\rangle$:
\begin{align}
\frac{\langle Y^{0}_{4}\rangle}{vol}=&\nu^{}_4 \left\langle Y^{0}_{4} \sum_{i=1}^3 (\phi_i)^4\right\rangle_0+ \nu^{}_6 \left\langle Y^{0}_{4}  (\phi_1\phi_2\phi_3)^2\right\rangle_0+\nu_6' \left \langle Y^{0}_{4} (\sum_{i=1}^3 \phi_i^2)(\sum_{i=1}^3 \phi_i^4)\right\rangle_0\nonumber\\
=&\nu^{}_4 \left\langle \phi^4 \right\rangle_0 \left\{Y^{0}_{4} f(\theta,\psi)\right\}_{S^2}+\nu^{}_6 \left\langle \phi^6 \right\rangle_0 \left\{Y^{0}_{4}  g(\theta,\psi) \right\}_{S^2}
+\nu_6' \left\langle \phi^6 \right\rangle_0 \left\{Y^{0}_{4}  f(\theta,\psi)\right\}_{S^2}.\nonumber\\
=& -\frac{1}{15\sqrt{\pi}}\nu^{}_4\langle\phi^4\rangle_0-\frac{1}{15\sqrt{\pi}}\nu_6'\langle\phi^6\rangle_0-\frac{1}{330\sqrt{\pi}}\nu^{}_6\langle\phi^6\rangle_0,
\end{align}
with
\begin{align}
f(\theta,\psi)=\cos(\theta)^4+ \sin(\theta)^4 \cos (\psi)^4+  \sin(\theta)^4 \sin (\psi)^4, \quad g(\theta,\psi)=\cos(\theta)^2 \sin(\theta)^2 \cos (\psi)^2 \sin(\theta)^2 \sin (\psi)^2.
\end{align}
The bracket $\{ A \}_{S^2}\equiv\int A \sin(\theta) \mathrm{d}\theta\, \mathrm{d}\psi /(4 \pi)$ is the average over the unit two-sphere. In the above derivation, we have used the parametrization $\phi_1$\,$=$\,$\phi \cos(\theta)$, $\phi_2$\,$=$\,$\phi \sin(\theta)\cos(\psi)$ and $\phi_3$\,$=$\,$\phi \sin(\theta)\sin(\psi)$.  We have also used the approximation $\langle\phi^n f(\theta,\psi)\rangle_0=\langle\phi^n\rangle_0 \{f(\theta,\psi)\}_{S_2}$. This crude approximation, strictly speaking, is valid in the semiclassical (approximate mean-field theory) region. The results in Fig.~2
of the main text, however, show that around the first-order phase transition at $V/t=0.05(5)$, these formulae yield reasonable qualitative results for the couplings $\nu_4,\nu_6$ and $\nu_6'$. In particular, $\nu_4$ changes sign at the phase transition as expected.

Through some simple algebra, the three anisotropy parameters can be expressed as
\begin{align}
\nu^{}_{4} &= -\frac{1}{vol}\frac{15\sqrt{\pi}}{\langle \phi^{4}\rangle}\left(\langle Y^{0}_{4}\rangle+\frac{\langle \phi^2 Y^{0}_{4}\rangle\langle \phi^4\rangle\langle \phi^6\rangle-\langle Y^{0}_{4}\rangle\langle\phi^6\rangle^2}{\langle \phi^6\rangle^2-\langle \phi^4\rangle\langle \phi^8\rangle}\right), \\
\nu^{}_{6} &= -\frac{1}{vol}\frac{231\sqrt{13\pi}}{\langle \phi^{6}\rangle}\langle Y^{0}_{6}\rangle,\\
\nu_6' &= \frac{1}{vol}15\sqrt{\pi}\left(\frac{\langle \phi^2 Y^{0}_{4}\rangle\langle \phi^4\rangle-\langle Y^{0}_{4}\rangle\langle\phi^6\rangle}{\langle\phi^6\rangle^2-\langle \phi^4\rangle\langle \phi^8\rangle}+\frac{7\sqrt{13}}{10\langle\phi^6\rangle}\langle Y^{0}_{6}\rangle\right).
\end{align}
We have also used the approximation $\langle \phi^n \rangle_0=\langle \phi^n \rangle$, which is valid since their difference is of linear order in the perturbation parameters $\nu_4,\nu_6$, or $\nu_6'$. These differences only introduce higher-order corrections in Eq.~\eqref{eq:seq9}.
\begin{figure}[b]
	\centering
	\includegraphics[width=0.8\columnwidth]{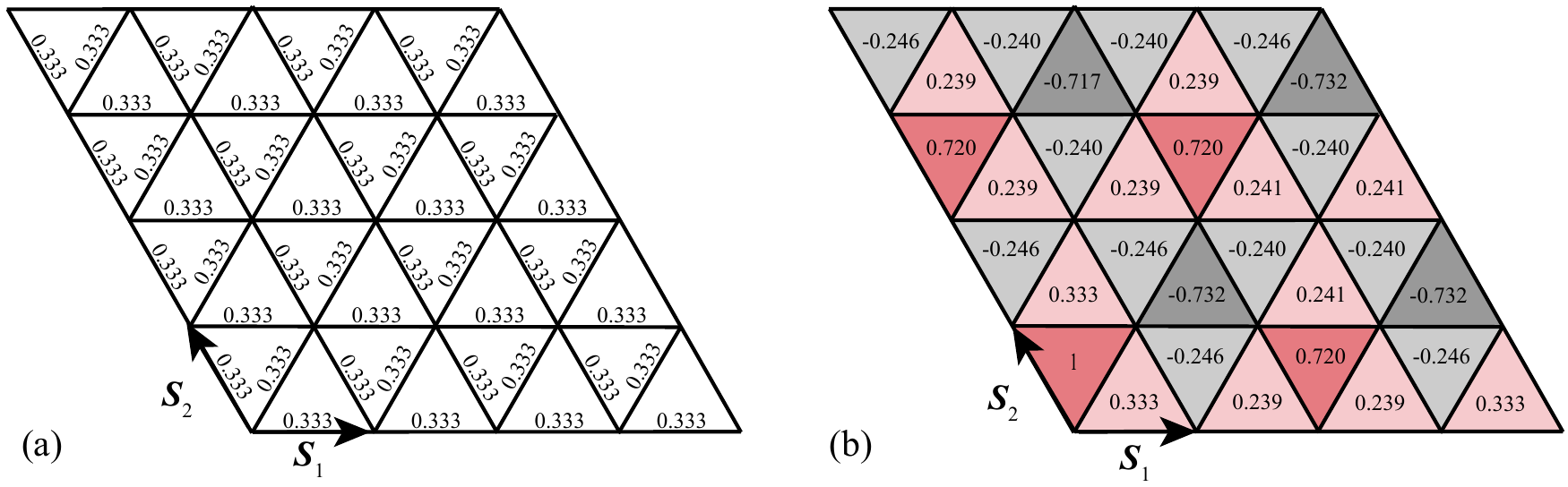}
	\caption{(a) Exact-diagonalization result for the real-space dimer density for the $L=4$ system at $V=0.3$; as previously, $\mathbf{s}_{1}$ and $\mathbf{s}_{2}$ are the primitive vectors. All bonds hold $1/3$ dimer density. (b) The real-space vison density, where we have set $v_1(0,0)=1$. Similar to the QMC results, four small triangles with the same sign of the vison density form bigger triangles, and the small triangles at the centers of the big ones have a larger absolute value of the density.}
	\label{fig:sfig4}
\end{figure}

\section*{SUPPLEMENTARY NOTE 6: Dimer density and correlation functions}
\label{sec:f}
In this section, we apply the Lanczos exact diagonalization (ED) method to a $4 \times 4$ ($N_\mathrm{bond}=48$) lattice within the VP phase. The number of configurations in the constrained Hilbert space is $586,695$. Using the ground-state wavefunction, we analyze the configurations symmetry of eight points of cubic order parameters without statistical error as the following explanation. As shown in Fig.~\ref{fig:sfig2}, the eight points lie in the eight octants, so each configuration of the wavefunction can be classified into a certain octant, except the ones on the boundaries between different octants. Then---akin to what is done in QMC---we classify configurations according to related octants and average over the ones in the same class to obtain the dimer density of a certain cubic order parameter point. It is worth noting that two classes of opposite octants have opposite signs for vison configurations but their dimer configurations are the same. Thus, we have to average the dimer configurations of the two opposite octants. The result [Fig.\ref{fig:sfig4}(a)] is similar to that obtained with QMC: there is still no dimer order to within the numerical precision of ED. We found that for each link, the real-space dimer density is $1/3$.

Similarly, we average the ED vison configurations in each octant. The histogram of the order parameter and the real-space vison density obtained via ED are consistent with our QMC results. The real-space vison density is shown in Fig.~\ref{fig:sfig4}(b). All these results strongly support the QMC conclusions that there is indeed a hidden-order vison plaquette phase between the spin liquid and nematic phases.

\begin{figure}[!]
	\centering
	\includegraphics[width=0.6\columnwidth]{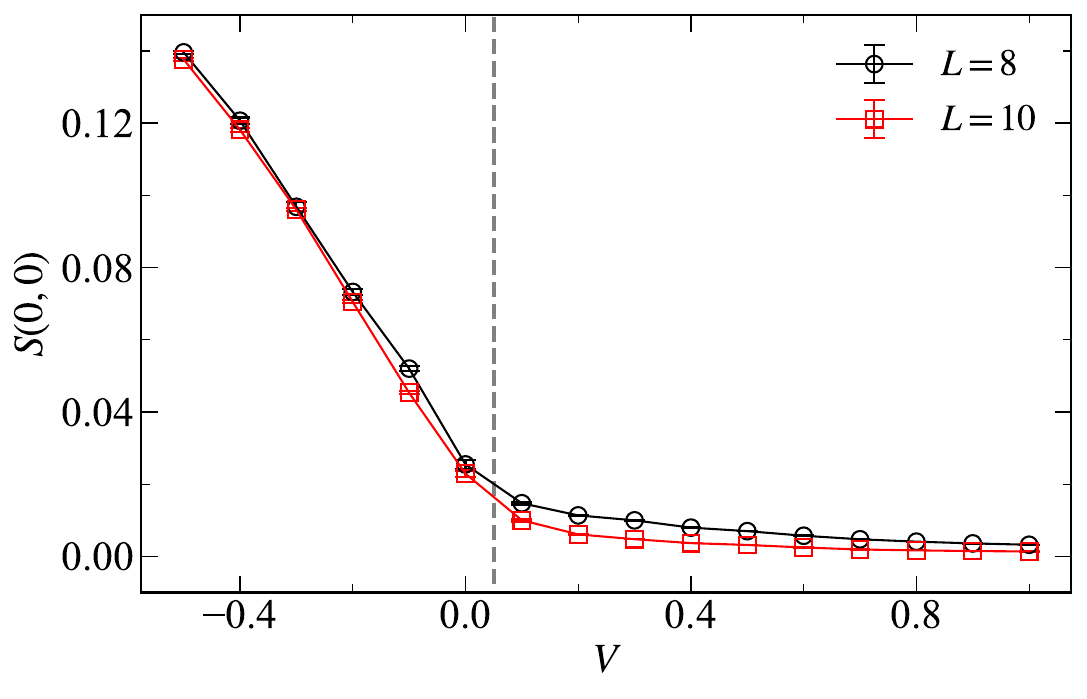}
	\caption{The structure factor of the dimer correlation functions at the $k=(0,0)$ point. This dimer order parameter vanishes near the transition point $V/t=0.05$ (gray dashed line), which indicates the transnational symmetry breaking at the first-order transition point near $V/t \sim 0$. But no sign for the continuous transition between VP and QSL phase. The error bars here represent the standard error of the mean, which is calculated as $\sigma/\sqrt{N}$, where $\sigma$ is one standard deviation and $N$ is the total number of independent samples.}
	\label{fig:dc}
\end{figure}

We also calculate the structure factor of the dimer-dimer correlation functions at the $\textbf{k}=(0,0)$ point as shown in Fig.~\ref{fig:dc}. The stucture factor is defined as
\begin{equation}
S(\textbf{k})=\frac{1}{N^2}\sum_{j=1}^{N}\sum_{k=1}^{N}e^{-i\textbf{k}(R_i-R_j)}[\langle D_{j}D_{k}\rangle - \langle D_{j}\rangle ^2],
\end{equation}
where $\langle D_{j}D_{k}\rangle = [(\langle d^1_{j}d^1_{k}\rangle + \langle d^2_{j}d^2_{k}\rangle + \langle d^3_{j}d^3_{k}\rangle]/3$ is the average dimer correlation functions of three orientation of rhombi. The dimer order parameter $S(0,0)$ is large in the LN phase, which preserves the transnational symmetry, and goes to zero at the first-order transition point between LN and VP phase, but it shows no sign for the continuous transition of the VP and QSL phase.


{\color{black}
\section*{SUPPLEMENTARY NOTE 7: Critical exponents of the continuous transition}
\label{sec:s}

The Binder ratio of the O(3) order parameter $\phi$ is defined as
\begin{equation}
B_2 \equiv\frac{\langle|\bm{\phi}|^4\rangle}{\langle|\bm{\phi}|^2\rangle^2},
\end{equation}
and the associated Binder cumulant $U_2$ is
\begin{equation}
U_2 \equiv\frac{5}{2}\left(1-\frac{3}{5}B_2\right).
\end{equation}
The scaling function for the vison order parameter $|\phi|$ and its Binder ratio $B_2(t,L)$ at reduced coupling $t=V-V_{\text{c}}$ and $t'=(V-V_{\text{c}})/V_{\text{c}}$ on
a lattice of length $L$ is
\begin{align}
\label{collapse1}
|\phi|L^{\beta/\nu}(t,L)=f(tL^{1/\nu}),\\
B_2(t',L)=f'(t'L^{1/\nu}),
\end{align}
where $V$ is the tunable parameter of the QLM in our work while $V_c$ is the critical point. This form can be used to carry out data collapse in a neighborhood of $t=0$, by graphing $|\phi|$ and $B_2(t,L)$ versus
$tL^{-1/\nu}$ and $t'L^{-1/\nu}$, adjusting $V_\text{c}$, $\beta$, and $\nu$ to obtain the tightest
collapse of the data onto a single curve~\cite{sandvikComputational2010}.

Setting $x=tL^{1/\nu}$ and $x'=t'L^{1/\nu}$,
we note that the both scaling functions $f(x)$ and $f'(x')$ are well-behaved and can be Taylor
expanded close to the critical point as
\begin{equation}
f(x)=q_0+q_1x+q_2x^2+q_3x^3+q_4x^4+\cdots.
\end{equation}
Although the properties of the cubic phase transition have been obtained from the effective field theory in the main text, in our analysis here, we keep the critical exponent $\nu$ and the critical point $V_c$ as two free parameters and sample them according to the goodness of fit. 

To obtained better estimates of $V_c$, $\beta$, and $\nu$ for the continuous VP--QSL phase transition in the QLM, we use a  polynomial fit for both the vison order parameter and its Binder ratio data and employ the bootstrap sampling method to perform their data collapse~\cite{wang2006high}. The vison order parameter and the Binder cumulant for different system sizes is shown in Figs.5(a) and (d) in the main text. The averages and error bars of the Binder ratio are calculated using bootstrap sampling, which resamples the  data of $\langle|\bm{\phi}|^4\rangle$ and $\langle|\bm{\phi}|^2\rangle$ by randomly sampling the original data~\cite{Newman1999}. Here, we use a polynomial fit to the third order for the raw vison order parameter and Binder ratio data, and the  $\chi^2/\mathrm{d.o.f}$ ($\chi^2$ per degree of freedom) for each fitting curve is close to 1, where $\chi^2=\sum_{i=1}^N(y'_i-y_i)^2/\sigma_i^2$, $y'_i$ is the data point on the fit curve, and $y_i$ and $\sigma_i$ are the raw data point and the error bar, respectively. For the Binder cumulant, we observe that the crossing between the curves for system sizes $L$ and $2L$  progressively shifts to the right because of the finite-size effect. Then, we set $V_c$, $\beta$, and $\nu$ as three free parameters to carry out the sampling for data collapse both for the vison order parameter and its Binder cumulant, and find the best fitting curve in the scaled $x=(V-V_{\text{c}})L^{1/\nu}$ and $x'=(V-V_{\text{c}})/V_{\text{c}}L^{1/\nu}$ with $\chi^2/\mathrm{d.o.f}\sim 1$. The three parameters $(V_c,\beta,\nu)$ are all selected randomly within an initial region 
according to our practical estimation. The converged distribution of the parameters $(V_c,\beta,\nu)$ is shown in Figs.5(b) and (e) in the main text, which compiles around 1000 independent polynomial fit results with $\chi^2/\mathrm{d.o.f}$ close to 1. To reduce the influence of the small system sizes, we only use $L\geq 8$ for the scaling. In this fashion, we determine $V_c=0.59(2)$, $\beta=0.33(5)$ and $\nu=0.75(8)$; the obtained $\beta$ and $\nu$ are consistent with the O(3) value of $\beta=0.3689(3)$ and $\nu=0.7112(5)$~\cite{Campostrini2002} within error bars. We then collapse the vison order parameter and its Binder ratio data with the obtained $V_c$, $\beta$, and $\nu$ and the results are shown in Figs.5(c) and (f) in the main text.

}

\end{widetext}

\end{document}